\documentclass[12pt,preprint]{aastex}

\usepackage{amssymb}
\usepackage{graphicx}
\usepackage{subfigure}

\shorttitle{The age comparison: SED Vs. CMD}
\shortauthors{Z. Fan, Z.-M. Li \& Gang Zhao}

\begin{document}

\title{The Ages of M31 Star Clusters: Spectral
  Energy Distribution Versus Color-Magnitude Diagram}

\author{Zhou Fan\altaffilmark{1}, Zhongmu Li\altaffilmark{2}, Gang
  Zhao\altaffilmark{1,3}} 

\altaffiltext{1}{Key Laboratory of Optical Astronomy, National
  Astronomical Observatories, Chinese Academy of Sciences, Beijing
  100101, China}  

\altaffiltext{2}{Dali College, Dali, Yunan 671000, China}

\altaffiltext{3}{School of Astronomy and Space Science, University of
  Chinese Academy of Sciences, Beijing 100049, China} 

\email{zfan@bao.ac.cn}

\begin{abstract}

  It is well-known that fitting the Color Magnitude Diagrams
  (CMDs) to the theoretical isochrones is the 
  main method to determine star cluster ages. However, when 
  the CMDs are not available, the Spectral Energy Distribution
  (SED)-fitting technique is the only other approach, although it
  suffers the age-metallicity-reddening degeneracy.
  In this work, we gather the ages, metallicities and masses
  of dozens of M31 star clusters from the CMD-fitting with HST images
  from the literatures for comparison. We check the reliability of the
  SED-fitting results with different models, i.e., Bruzual \& Charlot 2003
  model (BC03), Galaxy Evolutionary Synthesis Models ({\sl GALEV}) and
  Advanced Stellar Population Synthesis (ASPS) for the simple stellar 
populations (SSPs) with single stars (ss)-SSP/binary star (bs)-SSPs 
models. The photometry bands includes the Galaxy Evolution 
  Explorer {\sl GALEX} FUV/NUV bands, optical/near-infrared
  $UBVRIJHK$ bands, as well as the Wide-field Infrared Survey Explorer
  ({\sl WISE}) $W1$/$W2$ bands. The comparisons show that the
  SED-fitting ages agree well with the CMD-fitting ages, either with
  the fixed metallicity or with the free metallicity for both the BC03
  and the {\sl GALEV} model. However, for the ASPS models, it seems
  that SED-fitting results are systematic older than the CMD ages,
  especially for the ages log t $<9.0$ (yr). The fitting also shows
  that the {\sl GALEX} FUV/NUV-band are more important than the
  {\sl WISE} $W1$/$W2$  for constraining the ages, which 
  confirms the previous findings.  We also derived the masses of our
  sample star clusters from the BC03 and  {\sl GALEV} models and it is
  found that the values agree well with that in the literature. 
\end{abstract}

\keywords{galaxies: individual (M31) --- galaxies: star clusters ---
  globular clusters: general --- star clusters: general}

\section{Introduction}
\label{intro.sec}

Stellar population fitting techniques are important to determine
the physical parameters (e.g., ages, metallicities, and
masses) for the spatially unresolved stellar population. 
The CMD-fitting is the main approach to limit the
age of star clusters. It is reliable as the stars of a cluster
distribute on an isochrone if they have the same metallicity and
age. Because M31 is close enough to the Milky Way, the CMDs of M31
star clusters can be observed well via the Hubble 
Space Telescope (HST) and some large ground-based telescopes
\citep{rich05}. This makes us able to study these clusters using 
CMDs. For instance, \citet{p09a} observed the
star cluster VdB0 with the Wide Field  and Planetary Camera 2 (WFPC2)
on the HST and fit the observed CMD with the isochrone, giving the age,
metallicity and mass estimates. \citet{p09b} analyzed the number of M31
globular clusters (GCs) with reliable CMDs, and found that the number
of young and old GCs are 11 and 44, making use of the CMD structure
features. \citet{p10} obtained the ages and masses estimates of 19
clusters in M31 with the WFPC2 on HST data by fitting the CMD and the
theoretical isochrones. A sample of 24 star clusters with CMD-fitting
ages have been gathered. In addition, \citet{p11} present deep $BV$
photometry for four intermediate-age (1-9 Gyrs) M31 GCs with the wide field
channel of the Advanced Camera for Surveys (ACS) on board $HST$ and test how
the age estimates from integrated spectroscopy or photometry can be
supported by CMD-fitting. The authors also give the minimum ages
  of three M31 star clusters by fitting the isochrone and upper main
sequence (MS) stars and sub-giant branch (SGB) stars of
CMD. \citet{wh01} also determined the age of four M31 star clusters by
the isochrone fitting by eye and the software package MATCH, and it is
found that the ages from the two methods are in good agreement.  We are
showed that only the ages of a part of clusters can be estimated
similarly by different methods. Such works can be completed using some
classical stellar population models, e.g., BC03 \citep{bc03} and ASPS
\citep{lh08,li12,li17}. For the study of CMDs with different
  stellar population models one may see the papers
  \citep[e.g.,][]{li15,li17,ld18,ll18}.

  However, in some case, there is an age difference between SED and
  CMD fitting determinations, which is possibly caused by various
  constraints used in studies. In detail, the CMD ages are determined
  mainly by comparing the observed turn-off point to that of
  theoretical isochrones using an eye-aided method, when taking fixed
  distance and reddenning \citep{p09a,p10,p11}. This kind of
  isochrone-fit result is affected by IMF slightly, but it is not the
  case of SED fitting. In order to compare SED and CMD fitting
  determinations, both the CMD and luminosity function should be
  fitted, because luminosity function is also sensitive to IMF. It
  should be noted that there are obvious uncertainties in age
  determinations, even in CMD fitting. In the cases of some CMD parts
  or luminosity functions not fitted well, there will be large
  uncertainties in the CMD ages, which can contribute to the age
  difference. For instance, overshooting and stellar binarity can
  affect the result obviously. The age difference between CMD fitted
  ages from models can be as large as a factor of 0.6 \citep{p09a}.
  In addition, the use of different isochrones (considering the
  overshooting or not) also contributes to the age difference, which
  may lead to SED fitted ages younger than $\sim$ 0.4 dex compared to
  CMD fitted ages.

  The $\chi^2_{\rm min}$ SED-fitting of globular clusters is an
efficient method to determine the parameters on the basis of
multi-passband photometry/imaging data when the CMDs are not
available. Previously, \citet{rdg03} 
derive the ages, metallicities, and reddening of
the star clusters associated with NGC 3310 by the SED-fitting
method with the photometry of the ultraviolet (UV),
optical, and near-infrared (NIR) observations obtained with the {\sl
  Hubble Space Telescope} ({\sl HST}).
\citet{fan06,ma07,ma09,ma11,ma12,wang10,wang12} have done a 
series of SED-fitting works of M31 star clusters, based on the
Beijing--Arizona--Taiwan--Connecticut (BATC) multi-color photometry
system, with a 60/90cm Schmidt telescope. The simple stellar
population (SSP) models applied are the \citet[][henceforth BC03]{bc03}
and the Galaxy Evolutionary Synthesis Models \citep[{\sl
  GALEV};][]{lf06,kot09}. To achieve better results, the photometry on 
their bands, such as the broad-band $UBVRI$ filters, the Two Micron All
Sky Survey (2MASS) $JHK$ bands, the Galaxy Evolution Explorer ({\sl
  GALEX}) near-UV (NUV) and far-UV (FUV) channels, as well as the Sloan
Digital Sky Survey (SDSS) $ugriz$ bands have been applied. 
With considering the contributions of binary merger products in the
form of blue straggler stars \citet{fd12} also fitted the SEDs in
$UBVRIJHK$ and $ugriz$ bands of M31 globular-like 
clusters with the models including the bs-SSPs model. Since the
Wide-Field Infrared Survey Explorer \citep[{\sl WISE};][]{wri}
provides mid-infrared (mid-IR) $W1$/$W2$-band data, of which the
limiting magnitude could match the SED of M31 star clusters, we
also extend the wavelength range of SED to these bands. 

In our work, we derived the ages and metallicities of our sample M31
star clusters with SED-fitting based on the photometry bands from the
{\sl GALEX} FUV/NUV bands to the broadband $UBVRI$bands and NIR $JHK$
and {\sl WISE} $W1$/$W2$ bands. All the sample star clusters already
have the ages and metallicities fitted with the CMDs in the literature. Thus we
compared our results of the SED-fitting and that from the CMD-fitting
results in the literatures. The organization of the paper is as follows. In Section
\ref{mod.sec} we introduce the models mainly applied and convolutions of the 
passbands. In Section \ref{fit.sec}, we describe the selection of sample and 
introduce the methods adopted for the fits. In Section \ref{res.sec} we 
provide the best-fitting results as well as comparisons of the ages and 
masses based on $\chi^2_{\rm min}$ fitting with various models. 
Finally, the summary, concluding remarks and the future works are given 
in Section \ref{sum.sec}.  

\section{The Models and the $\chi_{\rm min}^2$-fitting Method}
\label{mod.sec}

The BC03 synthesis models \citep{bc03} is one of the most
commonly-used models for the SSP fitting, which could
provides the spectra and SEDs for different parameters, e.g., age,
metallicity. The models includes the 1994 and 2000 Padova stellar
evolutionary tracks as well as the stellar initial mass functions 
(IMFs) of \citet{sal} and \citet{chab03} and the wavelength coverage
is 91 {\AA}-160$\mu$m. The models includes six metallicity 
values ($Z=0.0001$,  0.0004, 0.004, 0.008, 0.02, and 0.05) for Padova
1994 tracks, while another six metallicities ($Z=0.0004$,
0.001, 0.004, 0.008, 0.019, and 0.03) are included for the Padova
2000 tracks. For the age, there are 221 values 0-20 Gyr in unequally
spaced time steps could be used. In our work, the upper limit of age
is set to 13.8 Gyr, which is Planck's latest estimate of age of
  the universe \citep{planck}. In order to reduce the  
  intervals of metallicity space and obtain better results which
  seems ``more accurate'' or ``higher resolution'' as the grids of
  metallicity space are smaller 
  in the models and thus closer to the nearest SSP in our fitting, we
  interpolate the original metallicities
the models  to 51 metallicities with equal steps in logarithmic space
rather than the original only six values. 
Since the \citet{chab03} IMF and Padova 2000 track are more up-to-date, 
they have been adopted in our fits and comparisons.

The {\sl GALEV} models \citep{kot09} are also commonly used to
explore the 
abundance evolution of gas and the spectral evolution of the stellar
populations, e.g., star clusters or galaxies. It provides the stellar
evolutionary tracks and isochrones from the Padova and Geneva groups.
In this work, Padova evolutionary tracks are adopted. The simple
stellar populations (SSPs) models provide 5001 ages  
4 Myr—16 Gyr and 7 metallicities  ($Z=0.0001$, 0.0004, 0.001, 0.004,
0.008, 0.02, and 0.05). The upper limit of age is also set to age
  of the universe 13.8 Gyr from the new result of Planck as done in BC03
model for the reason of comparisons. Since the   
number of metallicity values in the model are not enough for the fit,  
we interpolate the metallicities to a grid of 51 values as done previously, 
which can yield better results than the basic model set. The model
spectra are in the wavelength coverage 90 {\AA}-160 $\mu$m, which is,
convenient for us to convolve the model spectra with the filter
transmission curves and obtain the model magnitudes. There are
  two commonly-used options for the IMF: \citet{sal} and
  \citet{kro}. Thus, in our work, \citet{kro} IMF is adopted for the
  fitting as it is more up-to-date and more reasonable than the
  \citet{sal} for the  {\sl GALEV} model.

The ASPS models which built both single-star simple stellar
populations (ss-SSPs) and the binary-star simple stellar populations
(bs-SSPs) models \citep{li12,li16} is developed from
the rapid population synthesis (RPS) model \citep{lh08}. The main
feature of ASPS models is that it takes into account single stars,
binary stars and rotating stars, simultaneously. Thus different stellar
population models including classical ss-SSPs and bs-SSPs can be used
to study star clusters. In ASPS models, some typical parameter
distributions (e.g., orbital eccentricity and mass ratio) and widely
used IMFs \citep[e.g.,][]{sal} are used for the stars of a population
and the rapid stellar evolutionary code of \citet{hpt00,htp02} is used
for computing stellar evolution. The effect of 
rotation on stellar evolution is also considered using some recent
results \citep[e.g.,][]{geo13}. For the standard models, eight
metallicities (Z =0.0001, 0.0003, 0.001, 0.004, 0.008, 0.01, 0.02 and
0.03) and 151 ages (0-15 Gyr with an interval of 0.1 Gyr) have been
taken. The fractions of binary stars and rotating stars can change
from zero to 1. 

For BC03 SSP models,  {\sl GALEV} models and the ASPS ss-SSPs/bs-SSPs
models, the theoretical spectra have been convolved into the AB
system with the transmission of filters in  the FUV, NUV, $UBVRIJHK$,
$W1$, and $W2$ bands \citep{jar11}. The AB  magnitudes of the models
are calculated in the formula \citep[see, e.g.,][]{fan06,ma07,ma09,ma11,ma12,wang10,wang12} below,
\begin{equation}
  m_{\rm AB}(t)=-2.5~{\log~\frac{\int_{\lambda_1}^{\lambda_2}{{\rm
          d}\lambda}~{\lambda}~F_{\lambda}(\lambda,t)~R(\lambda)}{\int_{\lambda_{1}}^{\lambda_{2}}{{\rm
          d} \lambda}~{\lambda}~R(\lambda)}}-48.60,
  \label{eq1}
\end{equation}
in the formula, $R(\lambda)$ is the transmission of the filters and
$F_{\lambda}(\lambda,t)$ is the theoretical model flux, which is a function of
wavelength ($\lambda$) and evolutionary time ($t$); $\lambda_1$ and
$\lambda_2$ are the short- and long-wavelength cut-offs of the
respective filters.

\section{Cluster Sample Selection and $\chi_{\rm min}^2$ Fitting}
\label{fit.sec}

In order to compare the ages and mass estimates from CMD-fitting and
that from the SED-fitting, we gather a sample which have the
CMD-fitting from $HST$ observations previously. The GCs of Our sample
are from the master catalog of \citet{p10} and \citet{p11}.
\citet{p10} estimated the ages and metallicities of 19 star clusters
with the CMD-fitting method. The data is from HST/WFPC2 survey of the
bright young star clusters of M31. In order to analyze, the ages and
metallicities of six star clusters from \citet{p09b} and four clusters from
\citet{wh01} also included. \citet{p09b} used the HST/ACS archive
  data and the solar abundance isochrone of \citet{gir02} while
  \citet{wh01} used observational data from WFPC2 imager aboard the
  HST and the isochrone of \citet{ber94}. Thus the master catalog
includes 29   
GCs. In addition, \citet{p11} gives the lower limits of ages 
for three star clusters, B292, B337 and B350 by fitting the isochrones
with the observed CMDs, especially for the SGB stars. Our final
  sample consists of 32 star clusters with ages and metallicities  
  already derived using the CMD fitting technique.

We collected photometry of M31 star clusters and candidates
from the UV to the mid-IR. The {\sl GALEX} FUV/NUV data, the $UBVRI$ 
broad-band data and the 2MASS $JHK$ data were obtained from the
Revised Bologna Catalog of M31 globular clusters and
candidates\footnote{http://www.bo.astro.it/M31/} \citep[RBC v.5;][]
{gall04,gall06,gall09}. The data consists of combined photometric
  observations, including the most updated ones. The FUV/NUV-band
photometry are actually from RBC v.5,  which is compose of data of
\citet{rey07,kang12}. The $UBVRI$-band photometry are from RBC v.5,
which is composed of optical photometry of dozens of various works,
including the updated works, \citep[e.g.,][]{fan10,pea10}. For the
$JHK$-band photometry, although RBC v.5  provides the photometry
mainly from 
\citet{gall04}), \citet{p10} also give the aperture photometry of the
$r=10{\arcsec}.0$ from 2MASS-6X-PSC catalog. In our work, we merge the two
catalogues and the photometry with smaller errors is adopted if both RBC and
\citet{p10} have the available $JHK$-band photometry. In fact most of
the photometry is adopted from \citet{gall04}, which is more precise.
For the {\sl WISE} photometry: the central wavelengths of $W1$, $W2$,
$W3$, and $W4$ bands are 3.4, 4.6, 12 and 22 $\mu$m. The  
limiting magnitudes (adopting a signal-to-noise ratio of 5 for
11-frame composite images) are 17.11, 15.66, 11.40, and 7.97 mag
\citep{wri}. Since the limiting magnitude of the $W3$ and $W4$ bands
are not sufficiently high for our analysis. The photometry and
associated uncertainties of {\sl WISE} $W1$ and $W2$ bands (in
the Vega system), is in fact from the work of \citet{fan17}.
For the BC03/{\sl GALEV}  model fit, we use bands {\sl GALEX}  FUV/NUV,
$UBVRIHJK$, {\sl WISE} $W1$/$W2$ magnitude currently. Thus we only
have 12 bands for the SED-fitting. 

In order to fit the multi-band photometry based on the
convolved SED models discussed in Section~\ref{mod.sec}, we then
converted all Vega-based photometry to the AB system using the
\citet{kuru} SEDs of Vega, which could be used to be convolved with
  the filter transmission to obtain the system offset between the Vega
  magnitude and AB magnitude. The SED of Vega is well-studied, thus the
  errors of the system offset is too small to be considered, compared to the
  observational errors and model errors. We fit the SEDs of the
  observational magnitude and model magnitude, which should be in the
  same system, either in AB or Vega. Here we use the AB system for the
  SED-fitting, thus all the magnitudes are converted to the AB
  system except the FUV/NUV as they have been in AB system
  previously. Since this process only changes the offset of two
  systems, it dose not affect errors, e.g., $\sigma_{M,i}$ in
  equation~\ref{eq2}, at all.   

Since the reddening affects SED fitting seriously,
we adopted reddening values for our sample star clusters from
Table 2 of \citet{p10}, which is also a combination of the values from
their own work and those from literature. $A_{\lambda}$ is 
estimated with equations of \citet{ccm89}, and a typical foreground Milky 
Way extinction law, $R_V =3.1$ is adopted. SEDs are fitted by using the 
equation following,
\begin{equation}
  \chi^2_{\rm min}={\rm
    min}\left[\sum_{i=1}^{12}\left({\frac{M_{\lambda_i}^{\rm
          obs}-M_{\lambda_i}^{\rm mod}(t,\rm [Z/H]}
      {\sigma_{M,i}}}\right)^2\right],
  \label{eq2}
\end{equation}
where $M_{\lambda_j}^{\rm mod}(t,\rm [Z/H])$ is the $i^{\rm th}$
model magnitudes with age $t$ and metallicity $\rm [Z/H]$, while 
$M_{\lambda_i}^{\rm obs}$ is the observed, reddening-corrected magnitude 
in the $i^{\rm th}$ band.

For the ASPS model fit, we use only the band  {\sl GALEX}  FUV/NUV,
$UBVRIHJK$ as the model dose not provide the {\sl WISE} $W1$/$W2$ band
magnitude currently. Thus the Equation~\ref{eq2} is applied using
  only ten photometric bands.

For the uncertainty estimate in the fits, we use the Equation~(\ref{eq4}),
\begin{equation}
  \sigma_{M,i}^{2}=\sigma_{{\rm obs},M,i}^{2}+\sigma_{{\rm mod},M,i}^{2},
  \label{eq4}
\end{equation}
where $\sigma_{M,i}$ represents the total magnitude uncertainty in the 
$i^{\rm th}$ band. The photometric uncertainties in the RBC v.5 magnitudes
are estimated as \citet{fan16}, i.e., 0.08 mag in $U$ band, 0.05 mag in 
$BVRI$ bands, 0.1 mag in $J$ band, and 0.2 mag in $HK$ bands \citep{gall04}. 
The photometric errors in the {\sl GALEX} filters are provided in the RBC; 
for the $W1$/$W2$ bands, the uncertainties are included in
Table~1 of \citet{fan17}.  For the model uncertainties, we adopted the
typical error 0.05 mag for BC03 and {\sl GALEV} SSP models, as done
previously by \citet{fan06,ma07,ma09, wang10,fd14}. 

For the comparison, we also did the SED-fitting with the fixed
metallicity from the \citet{p10,p11} with the three models. The
fitting results are listed in the Tables~\ref{t2.tab} --\ref{t7.tab}.

\section{Fit Results and Discussion}
\label{res.sec}

\begin{figure}
  \resizebox{\hsize}{!}{\rotatebox{0}{\includegraphics{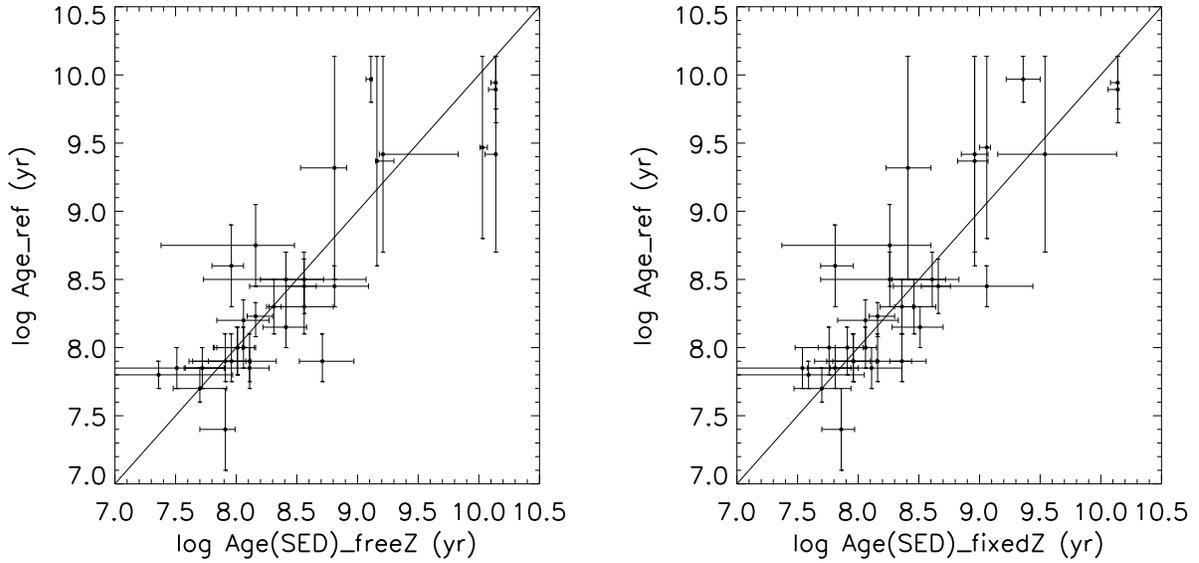}}}
  \caption{Age comparisons resulting from fits employing the BC03 models 
    using all-band photometry with free-Z (left) and those with fixed-Z (right).
    Padova 2000 evolutionary tracks and a \citet{chab03} IMF are adopted.}
  \label{fig1}
\end{figure}

To check the results from the SED-fitting, we compared with the
resulting ages from CMD-fitting from various sources in the literature.
Figure~\ref{fig1} shows the cluster ages as derived from SED-fitting with
Padova 2000 evolutionary tracks and a \citet{chab03} IMF, adopting
BC03 models, compared with age determinations from the CMD-fitting of 
\citet{p10,p11} in y-axis. In the left panel, ages in our work are
derived from SED-fitting with free-metallicity in the BC03 models. It is
found that the agreement is good overall except for a few outliers for
the older ones(log t $>$ $\sim10$) (yr). In the right panel, ages from our
estimates are derived with fixed metallicities, which are actually
from the CMD-fitting of literature \citet{p10,p11}. It is found that 
ages from free-metallicity is also consistent with the CMD results,
which is the same as that in the left panel. For the old part (log t
$>$ $\sim10$) (yr), the agreement seems slightly better than the left
panel. The comparison implies that the SED-fitting is reliable and
reasonable for the age fittings.  

\begin{figure}
  \resizebox{\hsize}{!}{\rotatebox{0}{\includegraphics{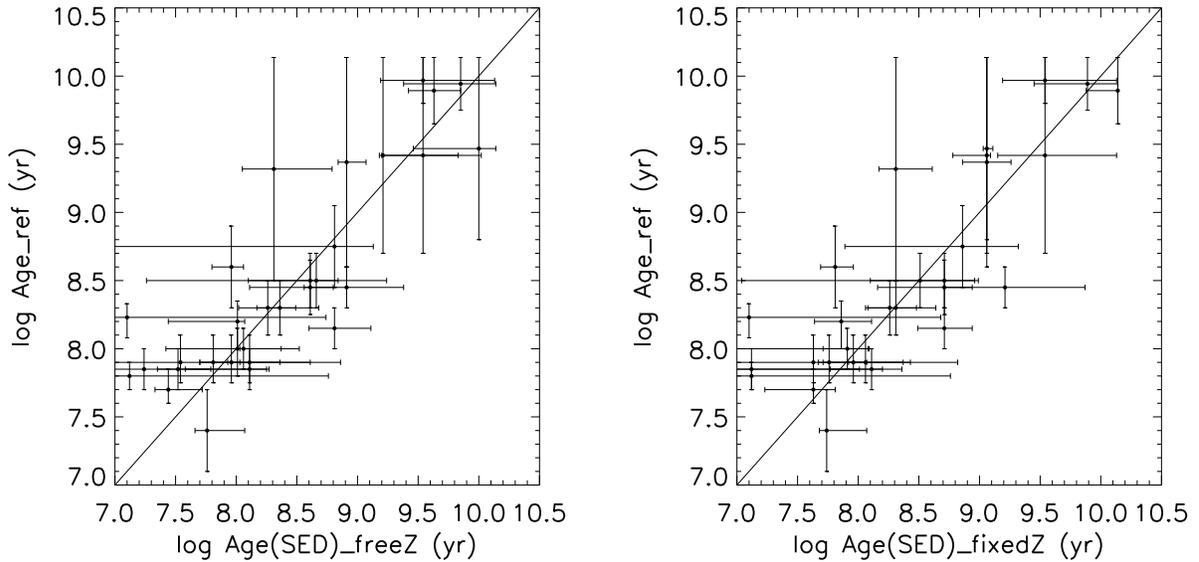}}}
  \caption{Same as Figure~\ref{fig1} but for the results fit with photometry
    without $FUV$/$NUV$ bands, which significantly affect the age
    estimate for the BC03 models.} 
  \label{fig2}
\end{figure}

\begin{figure}
  \resizebox{\hsize}{!}{\rotatebox{0}{\includegraphics{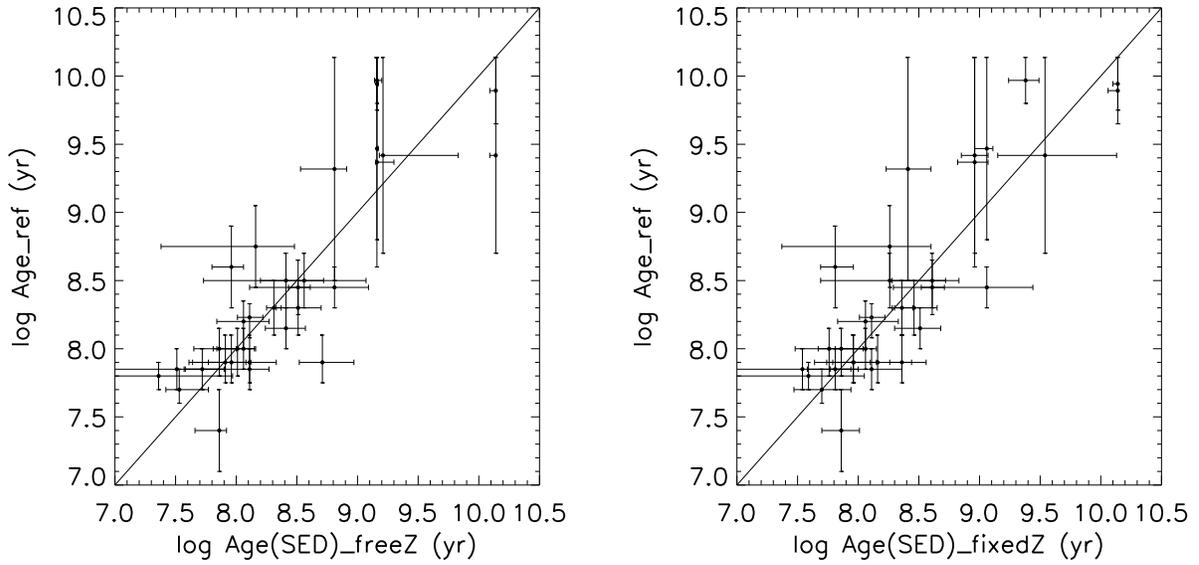}}}
  \caption{Same as Figure~\ref{fig2} but for the results fit without
    photometry only of WISE $W1$/$W2$ bands, which seems dose not affect
    much for the age estimate for the BC03 models.} 
  \label{fig3}
\end{figure}

To estimate the effects of using the  {\sl GALEX} FUV/NUV bands in the
SED-fitting, we have done the same comparison in Figure~\ref{fig2} as in
Figure~\ref{fig1}, but without the  {\sl GALEX} FUV/NUV bands. It can
be seen that the results are much worse than that in
Figure~\ref{fig1}, especially for the young clusters (log t $<8.5$
yr). The mean error increase from 0.22/0.24 dex to 0.41/0.40 dex in
  logarithmic space for the free-metallicity/fixed-metallicity
  fitting, compared to Figure~\ref{fig1}. The
same comparison is done in Figure~\ref{fig3} to estimate the 
effect of the WISE $W1$/$W2$ bands. However, it seems that the results  
do not change much as that in Figure~\ref{fig1},  indicating that the
WISE $W1$/$W2$ bands do not effect significantly, compared to that of the  {\sl
  GALEX} FUV/NUV bands. It confirms the conclusions of \citet{fan17},
i.e., the  {\sl GALEX} FUV/NUV bands effect much more
  significantly in the SED-fittings than the other bands, especially
  for the WISE $W1$/$W2$ bands. It is known that the
  integrated colours technique is affected by the effects of
  stochastic sampling of the IMF, while this could not be the case for
  CMD ages if they are derived from the best fitting isochrone, for
  example. Thus it is important to address the differences between
  the two different methods. It is discussed the stochastic fluctuations
  effect increases the uncertainties of predicted colours and
  magnitudes associated with the resulting model parameters,
  particularly for the lower mass clusters $ \le\rm 1\times 10^4~M_\odot$
  \citep{br10,fd12,ad13}. In our work, the fraction for the less
  massive clusters ($\le \rm 10^4~M_\odot$) is 5/28
  (from P10/P11) or 8/32 (for BC03 model with free-metallicity fitting)
  or 7/32  (for {\sl GALEV} model with free-metallicity fitting),
  which are small fraction. In fact the fraction is even less if we
  rule out the clusters whose mass are significant much smaller than $\le
  \rm 10^4~M_\odot$.  Thus we think the stochastic sampling effect is
  insignificant in our SED-fitting.

\begin{figure}
  \resizebox{\hsize}{!}{\rotatebox{0}{\includegraphics{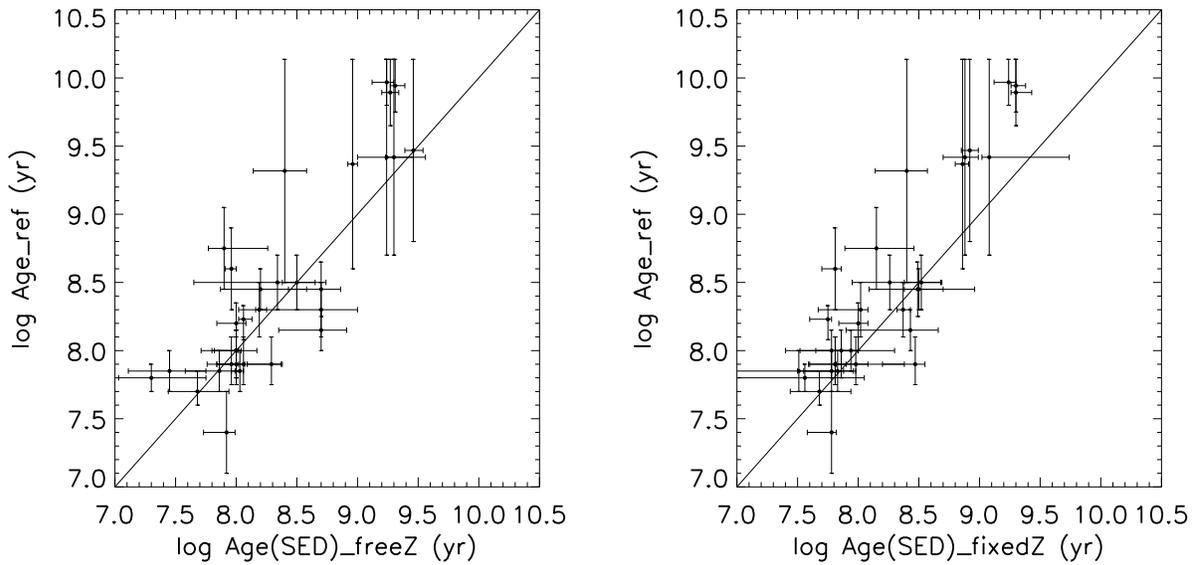}}}
  \caption{As Figure~\ref{fig1} but for {\sl GALEV} models and a \citet{kro}
    IMF with all-band photometry.}
  \label{fig4}
\end{figure}

In order to check the effect of different models, we also fit with
other models. Figure~\ref{fig4} is the same as Figure~\ref{fig1} but for
the {\sl GALEV} models with a \citet{kro} IMF. Similarly, we compare
the best-fitting SED resulting ages with free-metallicity are shown in the
left panel and those with fixed metallicity are shown in the right
panel.  It can be seen that the fitting results are systematically
younger than the CMD-fitting results for the older clusters, i.e.,
(log t $>9.5$ yr), although for the younger clusters the agreement is
well. This could be due to the model difference, which has been
mentioned by many previous works, e.g., \cite{fd12}.

\begin{figure}
  \resizebox{\hsize}{!}{\rotatebox{0}{\includegraphics{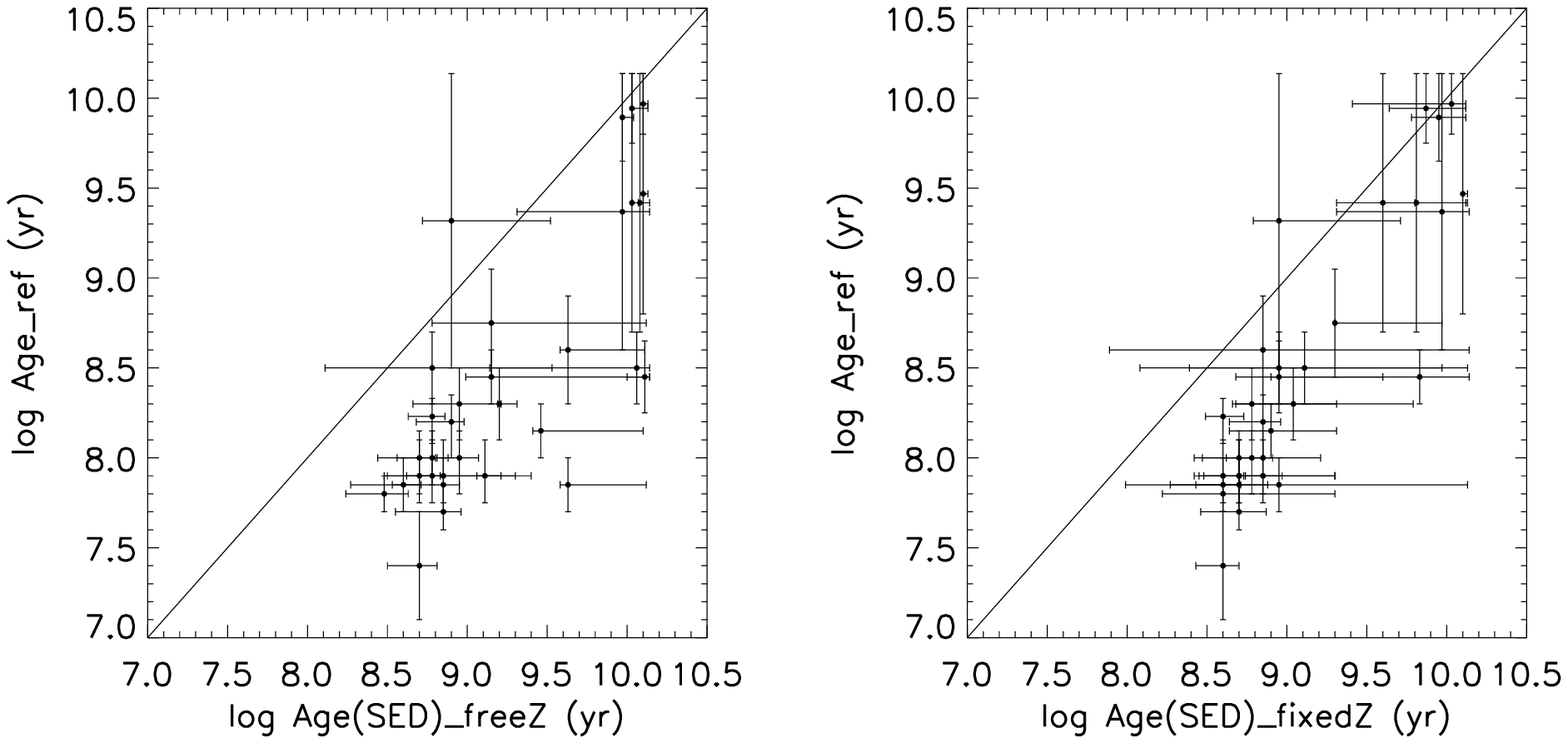}}}
  \caption{As Figure~\ref{fig1} but for the ASPS SSP models with
    all-band photometry without WISE W1/W2 band, which are not
    included in the model.}
  \label{fig5}
\end{figure}

\begin{figure}
  \resizebox{\hsize}{!}{\rotatebox{0}{\includegraphics{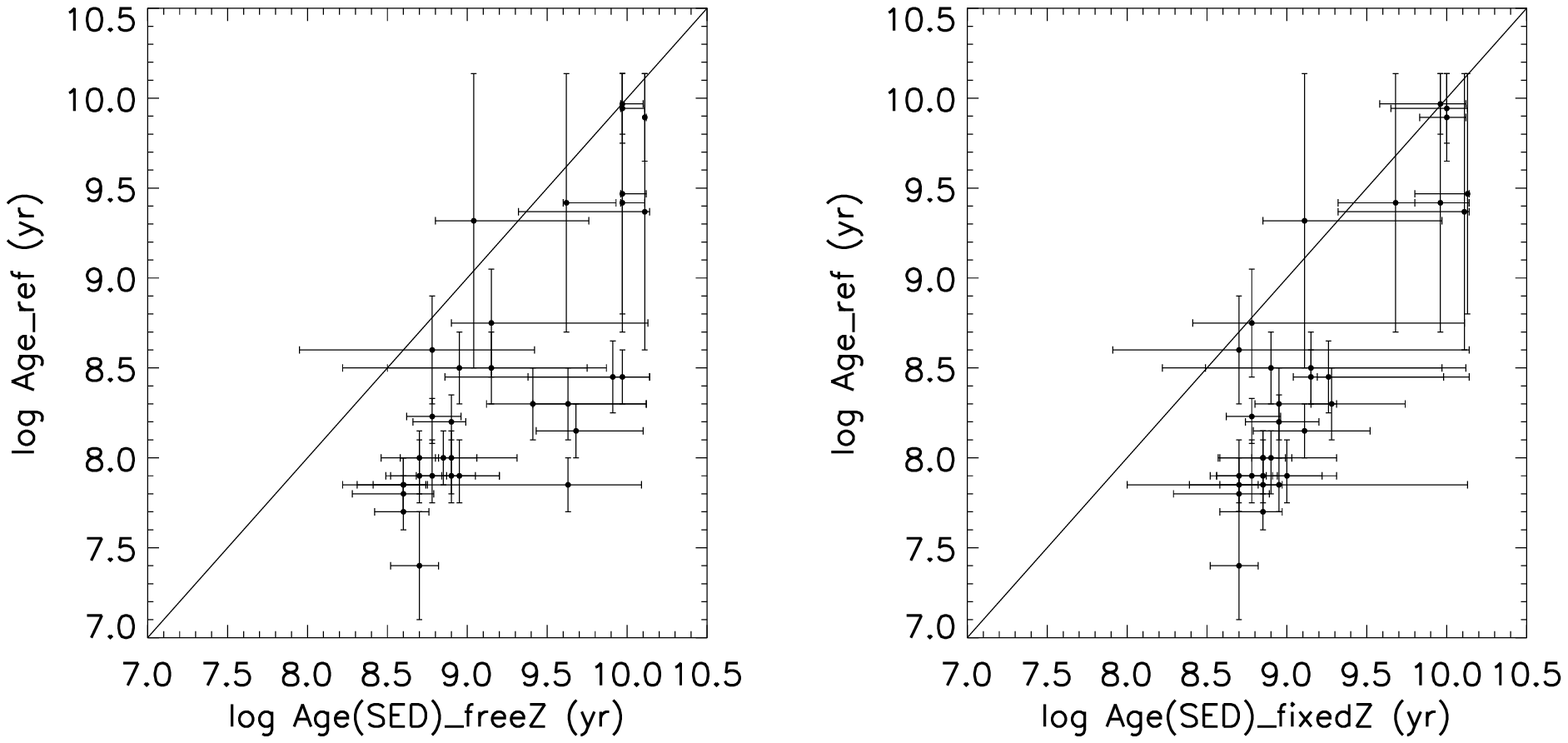}}}
  \caption{As Figure~\ref{fig1} but for the ASPS BSP models with
    all-band photometry without WISE W1/W2 band, which are not
    included in the model.}
  \label{fig6}
\end{figure}

Figures~\ref{fig5}--\ref{fig6} are the same as Figure~\ref{fig1} but for
the ASPS ss-SSPs/bs-SSPs models without WISE W1/W2 band photometry, as
it is not included in the the models.  We note the presence of systematic
differences in both left panel (metallicity-free fitting) and in the right
panel (metallicity-fixed fitting), in both of the ss-SSPs
(in Figure~\ref{fig5}) and bs-SSPs models (in Figure~\ref{fig6}).
We see that for the ASPS ss-SSPs models/bs-SSPs models, 
SED-fitting results are systematic older than the CMD ages ($\sim0.7$
dex), especially for the ages log t $<9.0$ (yr). It may be due to the
calibration of the models,  as there is no such bias for the
CMD-fitting and the SED-fitting with {\sl GALEV}  and BC03
models. It suggests that
there may be a problem with the adopted stellar evolution models
(independent from the inclusion of binaries). In the previous studies,
\citet{sal} and \citet{kro} IMFs were used and the RSG 
  star distribution was used to constrain stellar
  metallicity. As we see, the luminosity functions of many sample
  clusters were not fitted well, in particular for the bright end
  (e.g. B015D,B066, B040, B043 and B448). Some clusters are lack of
  reliable luminosity functions (B475 and V031) and their CMD ages
  were only from isochrone fitting, while some other clusters have no
  clear main sequences (e.g. B374, B222, B083, NB16 and B347 ) 
  and only lower limits were given for their ages. This should
  contribute to the age difference between CMD and SED
  determinations. 

  In CMD fittings, a BASTI isochrone database was used, which has
  taken into account
  overshooting\footnote{http://www.oa-teramo.inaf.it/BASTI/index.php}. 
  However, in the SED fittings, in particular ASPS fitting,
  overshooting usually was not considered. This may lead to some
  difference between SED and CMD fitted ages. If in SED fitting the
  age change caused by overshooting is similar to the case of CMD
  fitting ($\sim0.4$ dex) , it can roughly explain why the ages
  derived from ASPS models defer from CMD ages by 0.4--0.5 dex
  systematically. This effect is the same in both ss-SSP and bs-SSP
  models, so that we conclude that it is not due to the inclusion of
  binary stars. This effect is the same in both ss-SSP and bs-SSP
  models, so that we conclude that it is not due to the inclusion of
  binary stars. 

The masses of M31 star clusters are also important physical
  parameters. Firstly, we would like to briefly define the different ways in
  which the masses are obtained: 
  
  i. The masses calculated from the V0 magnitude of Table 2 of
  \citet{p10} and \citet{p11}.

  ii. The values from the literatures:  the masses provided by Table 2
  of \citet{p10}, which includes the results from \citet{p09b},
  \citet{wh01} and their own work, and \citet{p11}. 

  iii. The masses estimated from the SED-fitting directly in our work.
  
  Previously, \citet{p10} compute the masses of star clusters by using the
dereddened absolute $V_0$ magnitude, which are from RBC.

\begin{figure}
  \resizebox{\hsize}{!}{\rotatebox{0}{\includegraphics{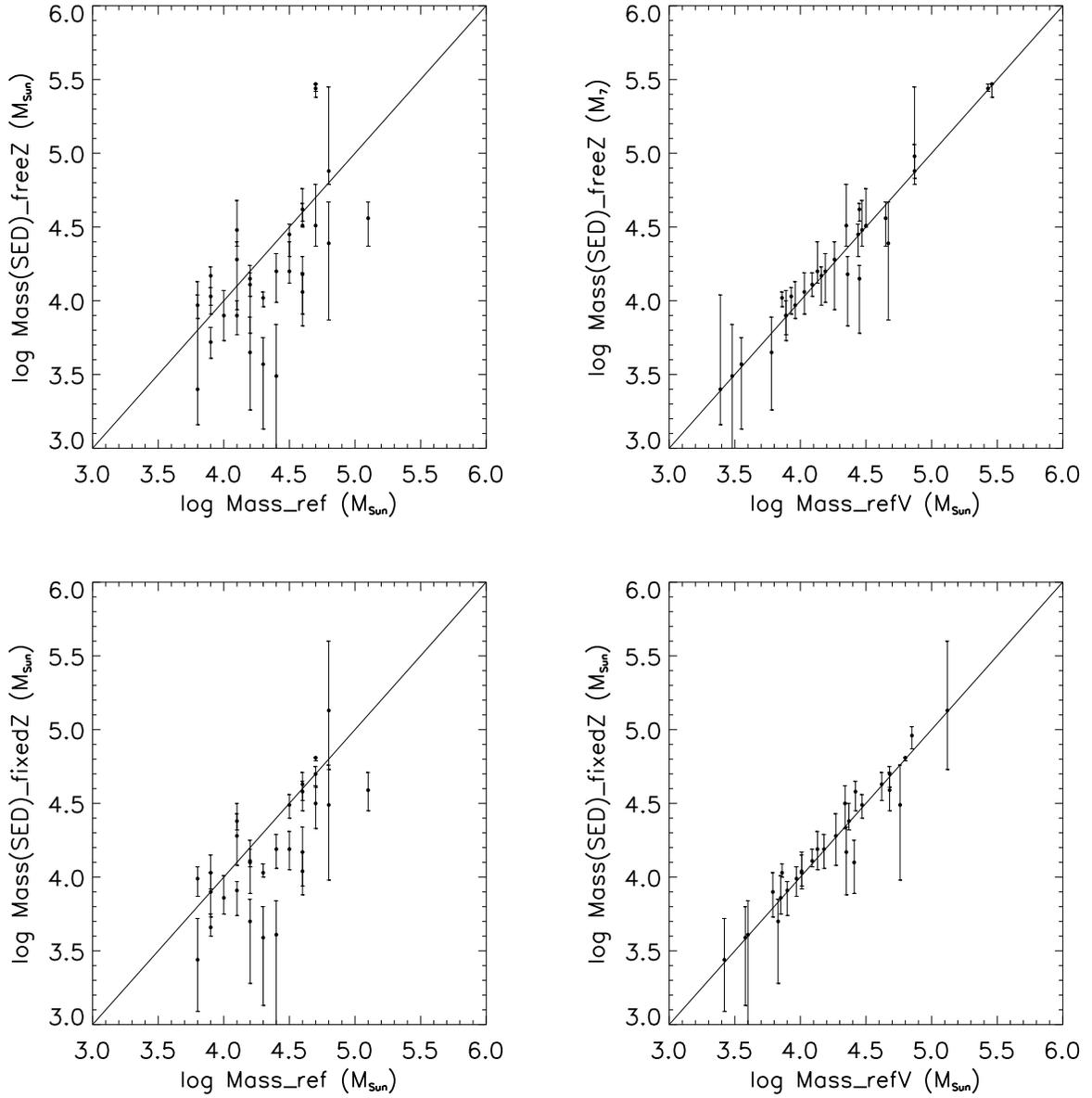}}}
  \caption{Mass comparisons resulting from fits employing the BC03 models
    using photometry with free-Z fit (top) and those with fixed-Z fit (bottom).
    Padova 2000 evolutionary tracks and a \citet{chab03} IMF are adopted.}
  \label{fig7}
\end{figure}

Since BC03 models provide the mass-to-light ratio in $V$-band for 
the given age and metallicity in their models, the masses can be
derived with the fitted parameters. In our work, the $V$-band
magnitude are from the updated version RBC v.5, which includes the 
updated photometry from \citet{fan10,pea10}. The distance module $\rm
(m-M)_0 = 24.47$ \citep{mc05} is adopted for the calculation of the
absolute magnitude. The extinction values are from the literatures
\citep[e.g.,][]{p10,p11}  and the extinction law is from
\citet{ccm89}. The comparison of our SED-fitting results (iii) and that from
literature (ii) is shown in figure~\ref{fig7}. The top panels are the
masses estimated with the free-metallicity SED-fitting 
results. In the left panel we use $V_0$ magnitudes from our work
  (iii) while in the right panel we adopt $V_0$ intrinsic magnitude from
  Table~2 of \citet{p10} and \citet{p11} (i).
It can be seen that the agreement is good for our estimates (left
  panel, iii) and even better if the $V_0$ magnitude of
\citet{p10,p11} are adopted (right panel, i). The bottom panels are the
fixed-metallicity SED-fitting results (iii). Similarly, the mass estimates
with intrinsic $V_0$ magnitude from Table~2 of \citet{p10} and
\citet{p11} (i) agrees with the the mass estimates of literature (ii) much
better than that with the $V_0$ magnitude in our SED-fitting (iii). This is
mainly due to the different $V_0$ magnitudes of Table~2 of \citet{p10}
and \citet{p11}, which used the old version of RBC and
that from our work, with the updated photometry in RBC.  It is also
found that the SED-fitting with/without fixed-metallically from
literature dose not change the agreement of the SED-fitting significantly. 

\begin{figure}
  \resizebox{\hsize}{!}{\rotatebox{0}{\includegraphics{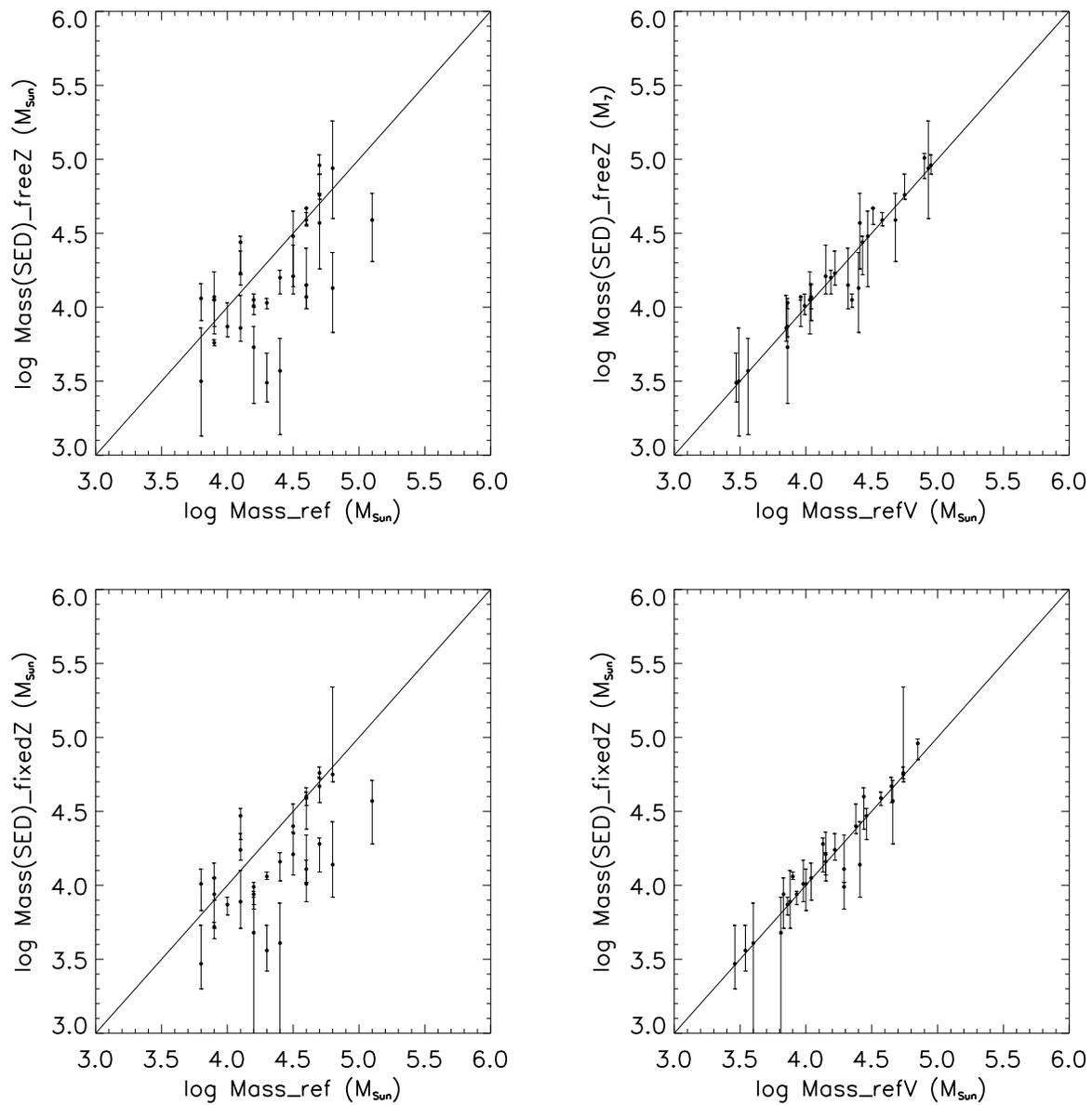}}}
  \caption{Same as Figure~\ref{fig7} but for {\sl GALEV} models and \citet{kro} 
    IMF using photometry with free-Z fit (top) and those with fixed-Z fit (bottom).}
  \label{fig8}
\end{figure}

Figure~\ref{fig8} is the same as figure~\ref{fig7} but for
mass-to-light ratios provided by the {\sl GALEV} models with a
\citet{kro} IMF. The conclusion is almost the same as that in
figure~\ref{fig7}, indicating that the masses calculated from 
the $V_0$  magnitude of Table~2 of \citet{p10} and \citet{p11} (in the
top right panel and bottom right panel, i) agrees with the values from
the literatures (ii) much better than that with the masses estimated from the
SED-fitting directly in our work  (in the top left panel and bottom
left panel, iii) . It is mainly due to the different $V_0$ magnitudes of
literatures and that in our work (the updated photometry are adopted),
although the mass-to-light ratios may be also slightly
different. Another finding is that there are some very massive
clusters (log $M/M_{\odot}\sim5.5$ for the metallicity-free models and
log $M/M_{\odot} \sim5.1$ for the metallicity-fixed models) predicted
in the BC03  models, but we do not find those counterparts in the {\sl GALEV}
models, which is supposed to be due to the difference of the models.

\section{Summary and Future Work}
\label{sum.sec}

We have collected photometric measurements of a sample of 32 star
clusters in M31, which have the SED-fitting results previously, from
the {\sl GALEX}  FUV/NUV to the  {\sl WISE}  mid-IR wavelength range to 
explore their importance for SED fits. We obtained the FUV, NUV,
$UBVRI$, and 2MASS $JHK$ data from the RBC v.5 catalog as well as the
work of \citet{p10,p11}. The {\sl WISE} $W1$/$W2$ band photometry was
from \citet{fan17}, which is actually downloaded from the IPAC/IRAS 
website. We applied the $\chi_{\rm min}^2$ technique for our SED-fitting and
all the upper limit of the age are set to 13.8 Gyr from Planck's
latest result. The reddening values are also from the work of \citet{p10,p11}.
 
We based our fits on three stellar population synthesis models.

1. The currently most up-to-date Padova 2000 evolutionary tracks and
the \citet{chab03} IMF, as implemented in the 
BC03 models. In general, our results agree well with 
previous determinations, either for the metallicity-free fitting
results or for the metallicity-fixed fitting results. In terms of the
cluster metallicities, our values agree well with those of
\citet{cbq}. Although most cluster ages of \citet{cw11} are upper
limits, their results agree reasonably well with ours.

2. We also compare the SED-fitting results derived from {\sl GALEV} models with
\citet{kro} IMF with the CMD-fitting results from literature.  As the
fitting for the BC03 models, our results agree well with previous
determinations, either for the metallicity-free fitting results or for
the metallicity-fixed fitting results. 

3. ASPS single-star (ss-SSPs) and the binary-star simple stellar
populations (bs-SSPs) model. Since the model dose not provide the
{\sl WISE} $W1$/$W2$ band, we do not use it for the SED-fitting. It is
found that a slight systematic difference between our fitting results
and previous CMD-fitting determinations, either for the
metallicity-free fitting results or for the metallicity-fixed fitting
results, in both ss-SSPs and bs-SSPs models. This could be an
  important results of the paper. In particular the fact that the ages
  based on ss-SSPs are also systematically different from CMD ages
  suggests that 
there may be a problem with the adopted stellar evolution models
(independent from the inclusion of binaries). Actually, the
``not-fitted-well'' of the luminosity function, lack of reliable luminosity
functions or clear main sequences, and only low limits also may lead
to the age difference. For instance, overshooting and stellar binarity
can affect the result obviously. The age difference between CMD fitted
ages from models can be as large as a factor of 0.6 dex. Besides, the use of
different isochrones (considered overshooting or not) also
contributes to the age difference up to 0.4 dex as well.

Therefore, our concluding remarks are:

1. The SED-fitting from the {\sl GALEX} FUV/NUV to the {\sl WISE}
$W1$/$W2$ band of our sample 32 star clusters in M31 agree well with
the previous CMD-fitting results, implying that the fitting method is
reliable and reasonable. 

2. The importance of the {\sl GALEX} FUV/NUV bands are much more
significant than the  {\sl WISE} $W1$/$W2$ bands for the SED-fitting,
which is consistent with the previous study, e.g., \citet{fan17},
  say, the  {\sl GALEX} FUV/NUV bands effect much more 
  significantly in the SED-fittings than the other bands, especially
  for the WISE $W1$/$W2$ bands.

3. It is found that the SED-fitting results with {\sl GALEV} models are
systematically younger than the CMD-fitting results for the older
clusters, i.e., (log t $>9.5$ yr), although for the younger clusters
the agreement is well. This could be due to the model difference,
which has been mentioned by many previous works, e.g., \cite{fd12}.

4. It is seems that for the ASPS ss-SSPs models/bs-SSPs models,
SED-fitting results are systematic older than the CMD ages ($\sim0.7$
dex), especially for the ages log t $<9.0$ (yr). It may be due to the
calibration of the models, as there is no such bias for the
CMD-fitting and the SED-fitting with  {\sl GALEV}  and BC03
models. In fact, age difference between CMD fitted ages from
different models can be as large as a factor of 0.6 and using different 
isochrones (e.g., considering the overshooting or not) could contributes to
the age difference $\sim0.4$ dex.  

In the future work, we would like to carry out systematic
  observations to give more constrains on the
  age/metallicity/reddening to disentangle the parameter
  degeneracy. Thus, it could be helpful if there is a photometric
  system which is sensitive to these physical parameters,
  e.g., age, metallicity, reddening.
The SAGE (Stellar Abundances and Galactic Evolution)
survey\footnote{http://sage.sagenaoc.science/~sagesurvey/}  
\citep[PI: Gang Zhao, see][]{fan18, zheng18a,zheng18b,zheng18c} apply
a brand-new SAGE photometry system, i.e., $u_{\rm SC}$,  $v_{\rm
  SAGE}$, gri, DDO51 and $H\alpha_n$/$H\alpha_w$ for the stellar
atmospheric parameters of 
$\sim500$ million stars. The self-designed $v_{\rm SAGE}$ band covers the
CaII H\&K lines, which is very sensitive to the metallicity of the
stellar populations, although it also effected by the age to some
  extent. The $H\alpha_n$/$H\alpha_w$ filters can be used to constrain the
  interstellar extinctions in M31, which is helpful to partly break the
  age-metallicity-reddening degeneracy. In the future work, we would
observe with the 
SAGE system obtained the $v_{\rm SAGE}$-band photometry of these star
clusters, which could improve the precision of the metallicity
estimates and helpful to determine the ages with the SED-fittings more
accurately. Further, we will enlarge this small M31 GC sample to all
the M31 GCs which have available photometry in all these bands (from
the  {\sl GALEX} FUV/NUV to the {\sl WISE}  $W1$/$W2$ as well as
the SAGE photometry) with the SED-fitting methods. Then the precise
ages of this large sample, which is comparable to the CMD-fitting
results, could be determined.

\acknowledgements

We thank the anonymous referee for his/her thorough review and
helpful comments and suggestions, which significantly contributed to
improving the manuscript.
This research has been supported by the National Program 
for Key Research and Development Project (grant 2016YFA0400804) and National
Key Basic Research Program of China (973 Program, grant 2015CB857002);
National Natural Science Foundation of China (NFSC) through grants 11390371,
11563002, U1631102, the Sino-German Center Project GZ 1284 and the
Youth Innovation Promotion Association, Chinese Academy of Sciences.    

\appendix		   


\clearpage
\pagestyle{empty}
\begin{deluxetable}{lrrrrrr}
  \tablecolumns{7} \tablewidth{0pc} \tablecaption{Ages of our sample
    star clusters  
    derived from SED fits of different passband combinations with BC03 models 
    and Padova 2000 stellar evolutionary tracks. The Chabrier (2003) IMF was 
    adopted. For the abundance of the Sun, $Z=0.019$ is applied for the
    models. The referential ages, metallicities and masses are from \citet{p10,p11}.
    \label{t2.tab}}
  \tablehead{\colhead{ID} &  \colhead{$\rm [Fe/H]$} &  \colhead{$\rm [Fe/H]_{P10/P11} $} &  \colhead{log $t$} &  \colhead{log $t\rm_{P10/P11}$} & \colhead{log $\rm M$} &  \colhead{log $\rm M_{P10/P11}$}   \\
    \colhead{} & \colhead{(dex)} &  \colhead{(dex)} & \colhead{(Gyr)} & \colhead{(Gyr)} &  \colhead{($M_\odot$)} & \colhead{($M_\odot$)}}
  \startdata
     B015D-D041 & $    0.28_{ -1.01}^{+  0.00} $ &   0.00 &  $    8.11_{ -0.53}^{+  0.16} $ & $    7.85_{ -0.15}^{+  0.15} $ & $  4.15_{ -0.37}^{+  0.09} $ &  4.20 \\
    B040-G102 & $    0.12_{ -0.78}^{+  0.16} $ &   0.00 &  $    7.96_{ -0.19}^{+  0.16} $ & $    7.90_{ -0.15}^{+  0.20} $ & $  4.06_{ -0.15}^{+  0.13} $ &  4.60 \\
    B043-G106 & $    0.15_{ -1.09}^{+  0.13} $ &   0.00 &  $    7.91_{ -0.27}^{+  0.17} $ & $    7.90_{ -0.15}^{+  0.20} $ & $  4.20_{ -0.21}^{+  0.12} $ &  4.40 \\
    B066-G128 & $    0.28_{ -1.38}^{+  0.00} $ &   0.00 &  $    7.51_{ -0.52}^{+  0.39} $ & $    7.85_{ -0.15}^{+  0.15} $ & $  3.65_{ -0.39}^{+  0.24} $ &  4.20 \\
    B081-G142 & $    0.28_{ -0.51}^{+  0.00} $ &   0.00 &  $    8.41_{ -0.19}^{+  0.17} $ & $    8.15_{ -0.15}^{+  0.15} $ & $  4.56_{ -0.19}^{+  0.11} $ &  5.10 \\
   B257D-D073 & $    0.28_{ -1.65}^{+  0.00} $ &   0.00 &  $    8.11_{ -0.50}^{+  0.22} $ & $    7.90_{ -0.15}^{+  0.20} $ & $  4.18_{ -0.35}^{+  0.12} $ &  4.60 \\
    B318-G042 & $    0.07_{ -0.99}^{+  0.21} $ &  -0.38 &  $    7.72_{ -0.15}^{+  0.19} $ & $    7.85_{ -0.15}^{+  0.15} $ & $  3.97_{ -0.09}^{+  0.16} $ &  3.80 \\
    B321-G046 & $    0.28_{ -0.30}^{+  0.00} $ &   0.68 &  $    8.16_{ -0.07}^{+  0.14} $ & $    8.23_{ -0.15}^{+  0.10} $ & $  4.11_{ -0.08}^{+  0.08} $ &  4.20 \\
    B327-G053 & $   -0.20_{ -0.76}^{+  0.46} $ &  -0.38 &  $    7.70_{ -0.22}^{+  0.22} $ & $    7.70_{ -0.10}^{+  0.15} $ & $  4.20_{ -0.08}^{+  0.20} $ &  4.50 \\
    B376-G309 & $   -0.06_{ -1.42}^{+  0.34} $ &   0.00 &  $    8.06_{ -0.22}^{+  0.09} $ & $    8.00_{ -0.15}^{+  0.15} $ & $  3.90_{ -0.13}^{+  0.10} $ &  4.10 \\
    B448-D035 & $   -1.63_{  0.00}^{+  0.85} $ &   0.00 &  $    8.71_{ -0.19}^{+  0.26} $ & $    7.90_{ -0.15}^{+  0.20} $ & $  4.48_{ -0.11}^{+  0.20} $ &  4.10 \\
    B475-V128 & $   -0.90_{ -0.73}^{+  0.83} $ &  -0.38 &  $    8.56_{ -0.29}^{+  0.24} $ & $    8.30_{ -0.20}^{+  0.20} $ & $  4.51_{ -0.14}^{+  0.28} $ &  4.70 \\
         V031 & $   -0.12_{ -1.17}^{+  0.40} $ &  -0.68 &  $    8.81_{ -0.70}^{+  0.28} $ & $    8.45_{ -0.15}^{+  0.15} $ & $  4.39_{ -0.52}^{+  0.28} $ &  4.80 \\
   VDB0-B195D & $   -0.07_{ -0.71}^{+  0.19} $ &   0.00 &  $    7.91_{ -0.21}^{+  0.08} $ & $    7.40_{ -0.30}^{+  0.30} $ & $  4.98_{ -0.15}^{+  0.08} $ & 99.99 \\
    B083-G146 & $   -1.63_{  0.00}^{+  0.13} $ &  -0.38 &  $   10.14_{ -0.09}^{+  0.00} $ & $    8.70_{  0.00}^{+  1.44} $ & $  5.47_{ -0.09}^{+  0.00} $ &  4.70 \\
    B222-G277 & $   -0.83_{ -0.18}^{+  0.19} $ &   0.00 &  $    9.16_{ -0.01}^{+  0.14} $ & $    8.60_{  0.00}^{+  1.54} $ & $  4.51_{ -0.01}^{+  0.25} $ &  4.60 \\
    B347-G154 & $   -1.63_{  0.00}^{+  0.13} $ &  -0.38 &  $   10.03_{ -0.02}^{+  0.04} $ & $    8.80_{  0.00}^{+  1.34} $ & $  5.44_{ -0.02}^{+  0.03} $ &  4.70 \\
    B374-G306 & $   -1.63_{  0.00}^{+  0.37} $ &   0.00 &  $    8.81_{ -0.28}^{+  0.10} $ & $    8.50_{  0.00}^{+  1.64} $ & $  4.17_{ -0.20}^{+  0.06} $ &  3.90 \\
         NB16 & $    0.28_{ -0.16}^{+  0.00} $ &   0.00 &  $    9.21_{ -0.03}^{+  0.62} $ & $    8.70_{  0.00}^{+  1.44} $ & $  4.88_{ -0.09}^{+  0.57} $ &  4.80 \\
    B049-G112 & $    0.27_{ -0.40}^{+  0.01} $ &   0.00 &  $    8.56_{ -0.15}^{+  0.10} $ & $    8.45_{ -0.20}^{+  0.20} $ & $  4.45_{ -0.15}^{+  0.07} $ &  4.50 \\
    B367-G292 & $    0.22_{ -0.17}^{+  0.06} $ &   0.00 &  $    8.31_{ -0.06}^{+  0.06} $ & $    8.30_{ -0.20}^{+  0.20} $ & $  4.02_{ -0.06}^{+  0.04} $ &  4.30 \\
    B458-D049 & $    0.28_{ -0.87}^{+  0.00} $ &   0.00 &  $    8.56_{ -0.36}^{+  0.16} $ & $    8.50_{ -0.20}^{+  0.20} $ & $  4.28_{ -0.34}^{+  0.12} $ &  4.10 \\
  B521-SK034A & $   -0.21_{ -0.75}^{+  0.16} $ &   0.00 &  $    7.96_{ -0.16}^{+  0.10} $ & $    8.60_{ -0.30}^{+  0.30} $ & $  3.72_{ -0.11}^{+  0.10} $ &  3.90 \\
         M039 & $   -1.23_{ -0.40}^{+  1.51} $ &   0.00 &  $    8.41_{ -0.68}^{+  0.66} $ & $    8.50_{ -0.20}^{+  0.20} $ & $  3.40_{ -0.24}^{+  0.64} $ &  3.80 \\
         M050 & $    0.28_{ -1.91}^{+  0.00} $ &   0.00 &  $    8.16_{ -0.78}^{+  0.32} $ & $    8.75_{ -0.30}^{+  0.30} $ & $  3.57_{ -0.44}^{+  0.18} $ &  4.30 \\
    B315-G038 & $   -1.39_{ -0.24}^{+  1.06} $ &  -0.38 &  $    8.01_{ -0.19}^{+  0.04} $ & $    8.00_{ -0.20}^{+  0.15} $ & $  4.62_{ -0.08}^{+  0.04} $ &  4.60 \\
    B319-G044 & $   -1.57_{ -0.06}^{+  0.88} $ &  -0.38 &  $    8.01_{ -0.20}^{+  0.15} $ & $    8.00_{ -0.20}^{+  0.15} $ & $  4.03_{ -0.12}^{+  0.06} $ &  3.90 \\
    B342-G094 & $   -0.09_{ -1.07}^{+  0.37} $ &  -0.38 &  $    8.06_{ -0.22}^{+  0.21} $ & $    8.20_{ -0.20}^{+  0.15} $ & $  3.90_{ -0.17}^{+  0.17} $ &  4.00 \\
    B368-G293 & $    0.28_{ -1.11}^{+  0.00} $ &   0.00 &  $    7.36_{ -0.50}^{+  0.61} $ & $    7.80_{ -0.10}^{+  0.10} $ & $  3.49_{ -0.55}^{+  0.35} $ &  4.40 \\
    B292-G010 & $   -0.91_{ -0.23}^{+  0.15} $ &  -1.80 &  $    9.11_{ -0.04}^{+  0.01} $ & $    9.80_{  0.00}^{+  0.34} $ & $  4.56_{ -0.02}^{+  0.05} $ & 99.99 \\
    B337-G068 & $   -1.45_{ -0.14}^{+  0.13} $ &  -1.28 &  $   10.14_{ -0.06}^{+  0.00} $ & $    9.65_{  0.00}^{+  0.49} $ & $  5.41_{ -0.05}^{+  0.00} $ & 99.99 \\
    B350-G162 & $   -1.63_{  0.00}^{+  0.08} $ &  -1.50 &  $   10.14_{ -0.04}^{+  0.00} $ & $    9.75_{  0.00}^{+  0.39} $ & $  5.49_{ -0.04}^{+  0.00} $ & 99.99 \\
 
  \enddata
\end{deluxetable}

\clearpage
\pagestyle{empty}
\begin{deluxetable}{lrrrrr}
  \tablecolumns{6} \tablewidth{0pc} \tablecaption{Same as Table~\ref{t2.tab}
    but for fixed-Z from literatures.
    \label{t3.tab}}
  \tablehead{\colhead{ID} &  \colhead{$\rm [Fe/H]_{P10/P11} $} &
    \colhead{log $t$} &  \colhead{log $t\rm_{P10/P11}$} & \colhead{log
      $\rm M$} &  \colhead{log $\rm M_{P10/P11}$} \\
    \colhead{} &  \colhead{(dex)} & \colhead{(Gyr)} & \colhead{(Gyr)} &  \colhead{($M_\odot$)} & \colhead{($M_\odot$)} }
  \startdata
     B015D-D041 &    0.00 &  $    8.11_{ -0.34}^{+  0.25} $ & $    7.85_{ -0.15}^{+  0.15} $ & $  4.10_{ -0.21}^{+  0.15} $ &  4.20 \\
    B040-G102 &    0.00 &  $    7.96_{ -0.17}^{+  0.19} $ & $    7.90_{ -0.15}^{+  0.20} $ & $  4.04_{ -0.10}^{+  0.13} $ &  4.60 \\
    B043-G106 &    0.00 &  $    7.96_{ -0.22}^{+  0.14} $ & $    7.90_{ -0.15}^{+  0.20} $ & $  4.19_{ -0.13}^{+  0.10} $ &  4.40 \\
    B066-G128 &    0.00 &  $    7.54_{ -0.56}^{+  0.40} $ & $    7.85_{ -0.15}^{+  0.15} $ & $  3.70_{ -0.42}^{+  0.15} $ &  4.20 \\
    B081-G142 &    0.00 &  $    8.51_{ -0.23}^{+  0.19} $ & $    8.15_{ -0.15}^{+  0.15} $ & $  4.59_{ -0.14}^{+  0.12} $ &  5.10 \\
   B257D-D073 &    0.00 &  $    8.16_{ -0.52}^{+  0.28} $ & $    7.90_{ -0.15}^{+  0.20} $ & $  4.17_{ -0.29}^{+  0.17} $ &  4.60 \\
    B318-G042 &   -0.38 &  $    7.81_{ -0.23}^{+  0.19} $ & $    7.85_{ -0.15}^{+  0.15} $ & $  3.99_{ -0.12}^{+  0.08} $ &  3.80 \\
    B321-G046 &    0.68 &  $    8.16_{ -0.07}^{+  0.14} $ & $    8.23_{ -0.15}^{+  0.10} $ & $  4.11_{ -0.04}^{+  0.08} $ &  4.20 \\
    B327-G053 &   -0.38 &  $    7.70_{ -0.23}^{+  0.24} $ & $    7.70_{ -0.10}^{+  0.15} $ & $  4.19_{ -0.14}^{+  0.12} $ &  4.50 \\
    B376-G309 &    0.00 &  $    8.06_{ -0.25}^{+  0.09} $ & $    8.00_{ -0.15}^{+  0.15} $ & $  3.91_{ -0.17}^{+  0.06} $ &  4.10 \\
    B448-D035 &    0.00 &  $    8.36_{ -0.10}^{+  0.20} $ & $    7.90_{ -0.15}^{+  0.20} $ & $  4.38_{ -0.06}^{+  0.12} $ &  4.10 \\
    B475-V128 &   -0.38 &  $    8.46_{ -0.28}^{+  0.18} $ & $    8.30_{ -0.20}^{+  0.20} $ & $  4.50_{ -0.17}^{+  0.12} $ &  4.70 \\
         V031 &   -0.68 &  $    9.06_{ -0.77}^{+  0.38} $ & $    8.45_{ -0.15}^{+  0.15} $ & $  4.49_{ -0.51}^{+  0.27} $ &  4.80 \\
   VDB0-B195D &    0.00 &  $    7.86_{ -0.16}^{+  0.11} $ & $    7.40_{ -0.30}^{+  0.30} $ & $  4.96_{ -0.09}^{+  0.06} $ & 99.99 \\
    B083-G146 &   -0.38 &  $    8.96_{ -0.11}^{+  0.11} $ & $    8.70_{  0.00}^{+  1.44} $ & $  4.70_{ -0.09}^{+  0.05} $ &  4.70 \\
    B222-G277 &    0.00 &  $    8.96_{ -0.14}^{+  0.11} $ & $    8.60_{  0.00}^{+  1.54} $ & $  4.63_{ -0.11}^{+  0.08} $ &  4.60 \\
    B347-G154 &   -0.38 &  $    9.06_{ -0.06}^{+  0.03} $ & $    8.80_{  0.00}^{+  1.34} $ & $  4.81_{ -0.02}^{+ -0.00} $ &  4.70 \\
    B374-G306 &    0.00 &  $    8.41_{ -0.18}^{+  0.19} $ & $    8.50_{  0.00}^{+  1.64} $ & $  4.03_{ -0.11}^{+  0.12} $ &  3.90 \\
         NB16 &    0.00 &  $    9.54_{ -0.39}^{+  0.59} $ & $    8.70_{  0.00}^{+  1.44} $ & $  5.13_{ -0.40}^{+  0.47} $ &  4.80 \\
    B049-G112 &    0.00 &  $    8.66_{ -0.14}^{+  0.10} $ & $    8.45_{ -0.20}^{+  0.20} $ & $  4.49_{ -0.09}^{+  0.07} $ &  4.50 \\
    B367-G292 &    0.00 &  $    8.36_{ -0.06}^{+  0.09} $ & $    8.30_{ -0.20}^{+  0.20} $ & $  4.03_{ -0.03}^{+  0.06} $ &  4.30 \\
    B458-D049 &    0.00 &  $    8.61_{ -0.33}^{+  0.22} $ & $    8.50_{ -0.20}^{+  0.20} $ & $  4.28_{ -0.20}^{+  0.15} $ &  4.10 \\
  B521-SK034A &    0.00 &  $    7.81_{ -0.12}^{+  0.15} $ & $    8.60_{ -0.30}^{+  0.30} $ & $  3.66_{ -0.06}^{+  0.09} $ &  3.90 \\
         M039 &    0.00 &  $    8.26_{ -0.57}^{+  0.46} $ & $    8.50_{ -0.20}^{+  0.20} $ & $  3.44_{ -0.35}^{+  0.28} $ &  3.80 \\
         M050 &    0.00 &  $    8.26_{ -0.89}^{+  0.34} $ & $    8.75_{ -0.30}^{+  0.30} $ & $  3.59_{ -0.46}^{+  0.21} $ &  4.30 \\
    B315-G038 &   -0.38 &  $    7.91_{ -0.24}^{+  0.15} $ & $    8.00_{ -0.20}^{+  0.15} $ & $  4.58_{ -0.13}^{+  0.07} $ &  4.60 \\
    B319-G044 &   -0.38 &  $    7.76_{ -0.28}^{+  0.28} $ & $    8.00_{ -0.20}^{+  0.15} $ & $  3.90_{ -0.17}^{+  0.13} $ &  3.90 \\
    B342-G094 &   -0.38 &  $    8.06_{ -0.23}^{+  0.27} $ & $    8.20_{ -0.20}^{+  0.15} $ & $  3.86_{ -0.11}^{+  0.15} $ &  4.00 \\
    B368-G293 &    0.00 &  $    7.59_{ -0.75}^{+  0.46} $ & $    7.80_{ -0.10}^{+  0.10} $ & $  3.61_{ -0.69}^{+  0.23} $ &  4.40 \\
    B292-G010 &   -1.80 &  $    9.36_{ -0.14}^{+  0.14} $ & $    9.80_{  0.00}^{+  0.34} $ & $  4.74_{ -0.08}^{+  0.10} $ & 99.99 \\
    B337-G068 &   -1.28 &  $   10.14_{ -0.08}^{+  0.00} $ & $    9.65_{  0.00}^{+  0.49} $ & $  5.42_{ -0.06}^{+  0.00} $ & 99.99 \\
    B350-G162 &   -1.50 &  $   10.14_{ -0.06}^{+  0.00} $ & $    9.75_{  0.00}^{+  0.39} $ & $  5.48_{ -0.05}^{+  0.00} $ & 99.99 \\

  \enddata
\end{deluxetable}

\clearpage
\pagestyle{empty}
\begin{deluxetable}{lrrrrrr}
  \tablecolumns{7} \tablewidth{0pc} \tablecaption{Same as Table~\ref{t2.tab}
    but for the {\sl GALEV} Models and a \citet{kro} IMF. $Z = 0.02$ (solar 
    metallicity) is adopted for the models.
    \label{t4.tab}}
  \tablehead{\colhead{ID} &  \colhead{$\rm [Fe/H]$} &  \colhead{$\rm [Fe/H]_{P10/P11} $} &  \colhead{log $t$} &  \colhead{log $t\rm_{P10/P11}$} & \colhead{log $\rm M$} &  \colhead{log $\rm M_{P10/P11}$}  \\
    \colhead{} & \colhead{(dex)} &  \colhead{(dex)} & \colhead{(Gyr)} & \colhead{(Gyr)} &  \colhead{($M_\odot$)} & \colhead{($M_\odot$)} }
  \startdata
     B015D-D041 & $   -0.35_{ -1.06}^{+  0.12} $ &  -0.02 &  $    8.03_{ -0.07}^{+  0.03} $ & $    7.85_{ -0.15}^{+  0.15} $ & $  4.05_{ -0.05}^{+  0.04} $ &  4.20 \\
    B040-G102 & $   -0.35_{ -0.72}^{+  0.21} $ &  -0.02 &  $    8.00_{ -0.16}^{+  0.07} $ & $    7.90_{ -0.15}^{+  0.20} $ & $  4.07_{ -0.08}^{+  0.08} $ &  4.60 \\
    B043-G106 & $   -0.38_{ -0.92}^{+  0.16} $ &  -0.02 &  $    7.96_{ -0.20}^{+  0.06} $ & $    7.90_{ -0.15}^{+  0.20} $ & $  4.20_{ -0.11}^{+  0.05} $ &  4.40 \\
    B066-G128 & $    0.40_{ -1.25}^{+  0.00} $ &  -0.02 &  $    7.45_{ -0.34}^{+  0.30} $ & $    7.85_{ -0.15}^{+  0.15} $ & $  3.73_{ -0.38}^{+  0.14} $ &  4.20 \\
    B081-G142 & $   -1.38_{ -0.32}^{+  0.66} $ &  -0.02 &  $    8.70_{ -0.35}^{+  0.21} $ & $    8.15_{ -0.15}^{+  0.15} $ & $  4.59_{ -0.28}^{+  0.18} $ &  5.10 \\
   B257D-D073 & $   -0.10_{ -1.07}^{+  0.50} $ &  -0.02 &  $    8.06_{ -0.22}^{+  0.32} $ & $    7.90_{ -0.15}^{+  0.20} $ & $  4.15_{ -0.16}^{+  0.25} $ &  4.60 \\
    B318-G042 & $   -0.42_{ -0.85}^{+  0.32} $ &  -0.40 &  $    7.86_{ -0.28}^{+  0.13} $ & $    7.85_{ -0.15}^{+  0.15} $ & $  4.06_{ -0.15}^{+  0.10} $ &  3.80 \\
    B321-G046 & $   -0.31_{ -0.56}^{+  0.22} $ &   0.65 &  $    8.06_{ -0.04}^{+  0.07} $ & $    8.23_{ -0.15}^{+  0.10} $ & $  4.01_{ -0.06}^{+  0.08} $ &  4.20 \\
    B327-G053 & $   -0.38_{ -0.69}^{+  0.38} $ &  -0.40 &  $    7.68_{ -0.24}^{+  0.26} $ & $    7.70_{ -0.10}^{+  0.15} $ & $  4.21_{ -0.12}^{+  0.21} $ &  4.50 \\
    B376-G309 & $   -0.38_{ -0.97}^{+  0.78} $ &  -0.02 &  $    8.00_{ -0.20}^{+  0.17} $ & $    8.00_{ -0.15}^{+  0.15} $ & $  3.86_{ -0.09}^{+  0.22} $ &  4.10 \\
    B448-D035 & $    0.40_{ -0.75}^{+  0.00} $ &  -0.02 &  $    8.29_{ -0.20}^{+  0.08} $ & $    7.90_{ -0.15}^{+  0.20} $ & $  4.44_{ -0.22}^{+  0.04} $ &  4.10 \\
    B475-V128 & $   -1.70_{  0.00}^{+  0.54} $ &  -0.40 &  $    8.70_{ -0.68}^{+  0.30} $ & $    8.30_{ -0.20}^{+  0.20} $ & $  4.57_{ -0.31}^{+  0.20} $ &  4.70 \\
         V031 & $    0.40_{ -1.33}^{+  0.00} $ &  -0.70 &  $    8.20_{ -0.33}^{+  0.38} $ & $    8.45_{ -0.15}^{+  0.15} $ & $  4.13_{ -0.30}^{+  0.24} $ &  4.80 \\
   VDB0-B195D & $   -0.38_{ -0.51}^{+  0.04} $ &  -0.02 &  $    7.92_{ -0.19}^{+  0.07} $ & $    7.40_{ -0.30}^{+  0.30} $ & $  5.01_{ -0.14}^{+  0.03} $ & 99.99 \\
    B083-G146 & $   -1.70_{  0.00}^{+  0.26} $ &  -0.40 &  $    9.30_{ -0.07}^{+  0.16} $ & $    8.70_{  0.00}^{+  1.44} $ & $  4.76_{ -0.03}^{+  0.14} $ &  4.70 \\
    B222-G277 & $   -0.31_{ -0.14}^{+  0.15} $ &  -0.02 &  $    8.96_{ -0.04}^{+  0.04} $ & $    8.60_{  0.00}^{+  1.54} $ & $  4.59_{ -0.04}^{+  0.05} $ &  4.60 \\
    B347-G154 & $   -1.63_{ -0.07}^{+  0.27} $ &  -0.40 &  $    9.46_{ -0.07}^{+  0.08} $ & $    8.80_{  0.00}^{+  1.34} $ & $  4.96_{ -0.06}^{+  0.07} $ &  4.70 \\
    B374-G306 & $   -0.06_{ -0.54}^{+  0.46} $ &  -0.02 &  $    8.40_{ -0.26}^{+  0.18} $ & $    8.50_{  0.00}^{+  1.64} $ & $  4.05_{ -0.23}^{+  0.19} $ &  3.90 \\
         NB16 & $    0.08_{ -0.50}^{+  0.19} $ &  -0.02 &  $    9.24_{ -0.24}^{+  0.32} $ & $    8.70_{  0.00}^{+  1.44} $ & $  4.94_{ -0.34}^{+  0.32} $ &  4.80 \\
    B049-G112 & $   -0.42_{ -1.27}^{+  0.35} $ &  -0.02 &  $    8.70_{ -0.27}^{+  0.16} $ & $    8.45_{ -0.20}^{+  0.20} $ & $  4.48_{ -0.34}^{+  0.17} $ &  4.50 \\
    B367-G292 & $    0.40_{ -0.11}^{+  0.00} $ &  -0.02 &  $    8.19_{ -0.03}^{+  0.06} $ & $    8.30_{ -0.20}^{+  0.20} $ & $  4.03_{ -0.04}^{+  0.03} $ &  4.30 \\
    B458-D049 & $    0.01_{ -0.15}^{+  0.24} $ &  -0.02 &  $    8.50_{ -0.12}^{+  0.15} $ & $    8.50_{ -0.20}^{+  0.20} $ & $  4.23_{ -0.08}^{+  0.15} $ &  4.10 \\
  B521-SK034A & $   -0.38_{ -0.57}^{+  0.05} $ &  -0.02 &  $    7.96_{ -0.05}^{+  0.04} $ & $    8.60_{ -0.30}^{+  0.30} $ & $  3.76_{ -0.02}^{+  0.02} $ &  3.90 \\
         M039 & $   -0.21_{ -1.49}^{+  0.60} $ &  -0.02 &  $    8.34_{ -0.69}^{+  0.40} $ & $    8.50_{ -0.20}^{+  0.20} $ & $  3.50_{ -0.37}^{+  0.36} $ &  3.80 \\
         M050 & $    0.40_{ -0.33}^{+  0.00} $ &  -0.02 &  $    7.90_{ -0.13}^{+  0.36} $ & $    8.75_{ -0.30}^{+  0.30} $ & $  3.49_{ -0.13}^{+  0.20} $ &  4.30 \\
    B315-G038 & $   -1.41_{ -0.28}^{+  0.57} $ &  -0.40 &  $    8.00_{ -0.18}^{+  0.02} $ & $    8.00_{ -0.20}^{+  0.15} $ & $  4.67_{ -0.11}^{+  0.00} $ &  4.60 \\
    B319-G044 & $   -1.56_{ -0.14}^{+  0.83} $ &  -0.40 &  $    8.00_{ -0.29}^{+  0.04} $ & $    8.00_{ -0.20}^{+  0.15} $ & $  4.07_{ -0.20}^{+ -0.02} $ &  3.90 \\
    B342-G094 & $   -0.38_{ -0.96}^{+  0.72} $ &  -0.40 &  $    8.00_{ -0.16}^{+  0.08} $ & $    8.20_{ -0.20}^{+  0.15} $ & $  3.87_{ -0.07}^{+  0.16} $ &  4.00 \\
    B368-G293 & $    0.40_{ -0.48}^{+  0.00} $ &  -0.02 &  $    7.30_{ -0.27}^{+  0.45} $ & $    7.80_{ -0.10}^{+  0.10} $ & $  3.57_{ -0.43}^{+  0.22} $ &  4.40 \\
    B292-G010 & $   -1.70_{  0.00}^{+  0.23} $ &  -1.82 &  $    9.24_{ -0.12}^{+  0.06} $ & $    9.80_{  0.00}^{+  0.34} $ & $  4.62_{ -0.12}^{+  0.04} $ & 99.99 \\
    B337-G068 & $   -1.09_{ -0.20}^{+  0.13} $ &  -1.30 &  $    9.27_{ -0.07}^{+  0.07} $ & $    9.65_{  0.00}^{+  0.49} $ & $  4.78_{ -0.05}^{+  0.06} $ & 99.99 \\
    B350-G162 & $   -1.63_{ -0.07}^{+  0.20} $ &  -1.52 &  $    9.31_{ -0.04}^{+  0.08} $ & $    9.75_{  0.00}^{+  0.39} $ & $  4.79_{ -0.02}^{+  0.08} $ & 99.99 \\

  \enddata
\end{deluxetable}

\clearpage
\pagestyle{empty}
\begin{deluxetable}{lrrrrr}
  \tablecolumns{6} \tablewidth{0pc} \tablecaption{Same as Table~\ref{t4.tab}
    but for fixed-Z from literatures.
    \label{t5.tab}}
  \tablehead{\colhead{ID} &  \colhead{$\rm [Fe/H]_{P10/P11} $} &  \colhead{log $t$} &  \colhead{log $t\rm_{P10/P11}$} & \colhead{log $\rm M$} &  \colhead{log $\rm M_{P10/P11}$} \\
    \colhead{} &  \colhead{(dex)} & \colhead{(Gyr)} & \colhead{(Gyr)} &  \colhead{($M_\odot$)} & \colhead{($M_\odot$)} }
  \startdata
     B015D-D041 &   -0.02 &  $    7.83_{ -0.28}^{+  0.05} $ & $    7.85_{ -0.15}^{+  0.15} $ & $  3.99_{ -0.15}^{+  0.03} $ &  4.20 \\
    B040-G102 &   -0.02 &  $    7.81_{ -0.21}^{+  0.27} $ & $    7.90_{ -0.15}^{+  0.20} $ & $  4.01_{ -0.12}^{+  0.16} $ &  4.60 \\
    B043-G106 &   -0.02 &  $    7.81_{ -0.22}^{+  0.09} $ & $    7.90_{ -0.15}^{+  0.20} $ & $  4.16_{ -0.13}^{+  0.06} $ &  4.40 \\
    B066-G128 &   -0.02 &  $    7.51_{ -0.63}^{+  0.45} $ & $    7.85_{ -0.15}^{+  0.15} $ & $  3.68_{ -0.72}^{+  0.24} $ &  4.20 \\
    B081-G142 &   -0.02 &  $    8.43_{ -0.53}^{+  0.23} $ & $    8.15_{ -0.15}^{+  0.15} $ & $  4.57_{ -0.29}^{+  0.14} $ &  5.10 \\
   B257D-D073 &   -0.02 &  $    7.98_{ -0.16}^{+  0.40} $ & $    7.90_{ -0.15}^{+  0.20} $ & $  4.11_{ -0.09}^{+  0.23} $ &  4.60 \\
    B318-G042 &   -0.40 &  $    7.78_{ -0.28}^{+  0.20} $ & $    7.85_{ -0.15}^{+  0.15} $ & $  4.01_{ -0.18}^{+  0.10} $ &  3.80 \\
    B321-G046 &    0.65 &  $    7.75_{ -0.15}^{+  0.03} $ & $    8.23_{ -0.15}^{+  0.10} $ & $  3.94_{ -0.07}^{+  0.02} $ &  4.20 \\
    B327-G053 &   -0.40 &  $    7.68_{ -0.24}^{+  0.26} $ & $    7.70_{ -0.10}^{+  0.15} $ & $  4.21_{ -0.14}^{+  0.15} $ &  4.50 \\
    B376-G309 &   -0.02 &  $    7.94_{ -0.29}^{+  0.36} $ & $    8.00_{ -0.15}^{+  0.15} $ & $  3.89_{ -0.18}^{+  0.21} $ &  4.10 \\
    B448-D035 &   -0.02 &  $    8.47_{ -0.27}^{+  0.08} $ & $    7.90_{ -0.15}^{+  0.20} $ & $  4.47_{ -0.16}^{+  0.05} $ &  4.10 \\
    B475-V128 &   -0.40 &  $    8.02_{ -0.35}^{+  0.06} $ & $    8.30_{ -0.20}^{+  0.20} $ & $  4.28_{ -0.19}^{+  0.04} $ &  4.70 \\
         V031 &   -0.70 &  $    8.50_{ -0.41}^{+  0.46} $ & $    8.45_{ -0.15}^{+  0.15} $ & $  4.14_{ -0.22}^{+  0.29} $ &  4.80 \\
   VDB0-B195D &   -0.02 &  $    7.78_{ -0.20}^{+  0.04} $ & $    7.40_{ -0.30}^{+  0.30} $ & $  4.96_{ -0.11}^{+  0.03} $ & 99.99 \\
    B083-G146 &   -0.40 &  $    8.88_{ -0.18}^{+  0.11} $ & $    8.70_{  0.00}^{+  1.44} $ & $  4.67_{ -0.11}^{+  0.06} $ &  4.70 \\
    B222-G277 &   -0.02 &  $    8.86_{ -0.06}^{+  0.05} $ & $    8.60_{  0.00}^{+  1.54} $ & $  4.59_{ -0.05}^{+  0.04} $ &  4.60 \\
    B347-G154 &   -0.40 &  $    8.92_{ -0.07}^{+  0.07} $ & $    8.80_{  0.00}^{+  1.34} $ & $  4.76_{ -0.04}^{+  0.04} $ &  4.70 \\
    B374-G306 &   -0.02 &  $    8.40_{ -0.26}^{+  0.17} $ & $    8.50_{  0.00}^{+  1.64} $ & $  4.05_{ -0.15}^{+  0.10} $ &  3.90 \\
         NB16 &   -0.02 &  $    9.08_{ -0.06}^{+  0.66} $ & $    8.70_{  0.00}^{+  1.44} $ & $  4.75_{ -0.05}^{+  0.59} $ &  4.80 \\
    B049-G112 &   -0.02 &  $    8.49_{ -0.11}^{+  0.21} $ & $    8.45_{ -0.20}^{+  0.20} $ & $  4.40_{ -0.05}^{+  0.15} $ &  4.50 \\
    B367-G292 &   -0.02 &  $    8.37_{ -0.05}^{+  0.06} $ & $    8.30_{ -0.20}^{+  0.20} $ & $  4.06_{ -0.02}^{+  0.03} $ &  4.30 \\
    B458-D049 &   -0.02 &  $    8.52_{ -0.14}^{+  0.16} $ & $    8.50_{ -0.20}^{+  0.20} $ & $  4.24_{ -0.07}^{+  0.11} $ &  4.10 \\
  B521-SK034A &   -0.02 &  $    7.81_{ -0.11}^{+  0.05} $ & $    8.60_{ -0.30}^{+  0.30} $ & $  3.72_{ -0.08}^{+  0.03} $ &  3.90 \\
         M039 &   -0.02 &  $    8.26_{ -0.31}^{+  0.43} $ & $    8.50_{ -0.20}^{+  0.20} $ & $  3.47_{ -0.17}^{+  0.26} $ &  3.80 \\
         M050 &   -0.02 &  $    8.15_{ -0.26}^{+  0.31} $ & $    8.75_{ -0.30}^{+  0.30} $ & $  3.56_{ -0.14}^{+  0.17} $ &  4.30 \\
    B315-G038 &   -0.40 &  $    7.86_{ -0.34}^{+  0.14} $ & $    8.00_{ -0.20}^{+  0.15} $ & $  4.60_{ -0.22}^{+  0.06} $ &  4.60 \\
    B319-G044 &   -0.40 &  $    7.78_{ -0.38}^{+  0.22} $ & $    8.00_{ -0.20}^{+  0.15} $ & $  3.94_{ -0.23}^{+  0.11} $ &  3.90 \\
    B342-G094 &   -0.40 &  $    8.00_{ -0.16}^{+  0.08} $ & $    8.20_{ -0.20}^{+  0.15} $ & $  3.87_{ -0.07}^{+  0.05} $ &  4.00 \\
    B368-G293 &   -0.02 &  $    7.56_{ -0.67}^{+  0.49} $ & $    7.80_{ -0.10}^{+  0.10} $ & $  3.61_{ -0.73}^{+  0.27} $ &  4.40 \\
    B292-G010 &   -1.82 &  $    9.24_{ -0.12}^{+  0.06} $ & $    9.80_{  0.00}^{+  0.34} $ & $  4.62_{ -0.12}^{+  0.02} $ & 99.99 \\
    B337-G068 &   -1.30 &  $    9.30_{ -0.04}^{+  0.13} $ & $    9.65_{  0.00}^{+  0.49} $ & $  4.75_{ -0.01}^{+  0.09} $ & 99.99 \\
    B350-G162 &   -1.52 &  $    9.30_{ -0.04}^{+  0.08} $ & $    9.75_{  0.00}^{+  0.39} $ & $  4.79_{ -0.01}^{+  0.06} $ & 99.99 \\

  \enddata
\end{deluxetable}

\clearpage
\pagestyle{empty}
\begin{deluxetable}{lrrrr}
  \tablecolumns{5} \tablewidth{0pc} \tablecaption{Same as Table~\ref{t2.tab}
    but for the ASPS bsp-SSPs Models. $Z = 0.02$ (solar metallicity) is adopted for the models.
    \label{t6.tab}}
  \tablehead{\colhead{ID} &  \colhead{$\rm [Fe/H]$} &  \colhead{$\rm [Fe/H]_{P10/P11} $} &  \colhead{log $t$} &  \colhead{log $t\rm_{P10}$}  \\
    \colhead{} & \colhead{(dex)} &  \colhead{(dex)} & \colhead{(Gyr)} & \colhead{(Gyr)} }
  \startdata
     B015D-D041 & $   -1.32_{ -0.50}^{+  0.94} $ &  -0.02 &  $    9.63_{ -1.41}^{+  0.46} $ & $    7.85_{ -0.15}^{+  0.15} $  \\
    B040-G102 & $   -0.32_{ -0.62}^{+  0.38} $ &  -0.02 &  $    8.90_{ -0.22}^{+  0.30} $ & $    7.90_{ -0.15}^{+  0.20} $  \\
    B043-G106 & $   -0.02_{ -0.46}^{+  0.20} $ &  -0.02 &  $    8.70_{ -0.18}^{+  0.17} $ & $    7.90_{ -0.15}^{+  0.20} $  \\
    B066-G128 & $    0.18_{ -0.78}^{+  0.00} $ &  -0.02 &  $    8.60_{ -0.29}^{+  0.15} $ & $    7.85_{ -0.15}^{+  0.15} $  \\
    B081-G142 & $   -1.42_{ -0.40}^{+  0.88} $ &  -0.02 &  $    9.68_{ -0.25}^{+  0.42} $ & $    8.15_{ -0.15}^{+  0.15} $  \\
   B257D-D073 & $    0.18_{ -0.67}^{+  0.00} $ &  -0.02 &  $    8.78_{ -0.29}^{+  0.27} $ & $    7.90_{ -0.15}^{+  0.20} $  \\
    B318-G042 & $    0.08_{ -0.59}^{+  0.10} $ &  -0.40 &  $    8.60_{ -0.19}^{+  0.14} $ & $    7.85_{ -0.15}^{+  0.15} $  \\
    B321-G046 & $    0.18_{ -0.35}^{+  0.00} $ &   0.65 &  $    8.78_{ -0.16}^{+  0.18} $ & $    8.23_{ -0.15}^{+  0.10} $  \\
    B327-G053 & $    0.08_{ -0.55}^{+  0.10} $ &  -0.40 &  $    8.60_{ -0.18}^{+  0.16} $ & $    7.70_{ -0.10}^{+  0.15} $  \\
    B376-G309 & $   -0.02_{ -0.69}^{+  0.20} $ &  -0.02 &  $    8.85_{ -0.27}^{+  0.46} $ & $    8.00_{ -0.15}^{+  0.15} $  \\
    B448-D035 & $    0.18_{ -0.33}^{+  0.00} $ &  -0.02 &  $    8.95_{ -0.11}^{+  0.25} $ & $    7.90_{ -0.15}^{+  0.20} $  \\
    B475-V128 & $   -1.12_{ -0.70}^{+  0.81} $ &  -0.40 &  $    9.63_{ -0.22}^{+  0.49} $ & $    8.30_{ -0.20}^{+  0.20} $  \\
         V031 & $   -0.02_{ -1.80}^{+  0.20} $ &  -0.70 &  $    9.97_{ -1.11}^{+  0.17} $ & $    8.45_{ -0.15}^{+  0.15} $  \\
   VDB0-B195D & $   -0.02_{ -0.49}^{+  0.20} $ &  -0.02 &  $    8.70_{ -0.18}^{+  0.12} $ & $    7.40_{ -0.30}^{+  0.30} $  \\
    B083-G146 & $   -0.12_{ -1.03}^{+  0.20} $ &  -0.40 &  $    9.97_{ -0.01}^{+  0.14} $ & $    8.70_{  0.00}^{+  1.44} $  \\
    B222-G277 & $   -0.02_{ -1.11}^{+  0.17} $ &  -0.02 &  $   10.11_{ -0.79}^{+  0.03} $ & $    8.60_{  0.00}^{+  1.54} $  \\
    B347-G154 & $   -0.12_{ -0.59}^{+  0.21} $ &  -0.40 &  $    9.97_{ -0.01}^{+  0.15} $ & $    8.80_{  0.00}^{+  1.34} $  \\
    B374-G306 & $    0.18_{ -0.58}^{+  0.00} $ &  -0.02 &  $    9.04_{ -0.24}^{+  0.72} $ & $    8.50_{  0.00}^{+  1.64} $  \\
         NB16 & $    0.18_{ -0.76}^{+  0.00} $ &  -0.02 &  $    9.62_{ -0.02}^{+  0.31} $ & $    8.70_{  0.00}^{+  1.44} $  \\
    B049-G112 & $   -1.22_{ -0.60}^{+  0.93} $ &  -0.02 &  $    9.91_{ -0.53}^{+  0.23} $ & $    8.45_{ -0.20}^{+  0.20} $  \\
    B367-G292 & $   -0.92_{ -0.31}^{+  0.45} $ &  -0.02 &  $    9.41_{ -0.29}^{+  0.71} $ & $    8.30_{ -0.20}^{+  0.20} $  \\
    B458-D049 & $    0.18_{ -0.79}^{+  0.00} $ &  -0.02 &  $    9.15_{ -0.65}^{+  0.72} $ & $    8.50_{ -0.20}^{+  0.20} $  \\
  B521-SK034A & $   -0.32_{ -1.50}^{+  0.50} $ &  -0.02 &  $    8.78_{ -0.83}^{+  0.64} $ & $    8.60_{ -0.30}^{+  0.30} $  \\
         M039 & $    0.18_{ -2.00}^{+  0.00} $ &  -0.02 &  $    8.95_{ -0.73}^{+  0.80} $ & $    8.50_{ -0.20}^{+  0.20} $  \\
         M050 & $   -0.92_{ -0.90}^{+  1.10} $ &  -0.02 &  $    9.15_{ -0.25}^{+  0.98} $ & $    8.75_{ -0.30}^{+  0.30} $  \\
    B315-G038 & $   -0.02_{ -0.63}^{+  0.20} $ &  -0.40 &  $    8.70_{ -0.24}^{+  0.12} $ & $    8.00_{ -0.20}^{+  0.15} $  \\
    B319-G044 & $   -0.52_{ -0.34}^{+  0.44} $ &  -0.40 &  $    8.90_{ -0.10}^{+  0.16} $ & $    8.00_{ -0.20}^{+  0.15} $  \\
    B342-G094 & $   -0.22_{ -0.64}^{+  0.30} $ &  -0.40 &  $    8.90_{ -0.24}^{+  0.09} $ & $    8.20_{ -0.20}^{+  0.15} $  \\
    B368-G293 & $    0.18_{ -1.18}^{+  0.00} $ &  -0.02 &  $    8.60_{ -0.32}^{+  0.19} $ & $    7.80_{ -0.10}^{+  0.10} $  \\
    B292-G010 & $   -0.02_{ -0.21}^{+  0.20} $ &  -1.82 &  $    9.97_{ -0.01}^{+  0.13} $ & $    9.80_{  0.00}^{+  0.34} $  \\
    B337-G068 & $   -0.12_{ -0.15}^{+  0.21} $ &  -1.30 &  $   10.11_{ -0.00}^{+  0.01} $ & $    9.65_{  0.00}^{+  0.49} $  \\
    B350-G162 & $   -0.12_{ -0.15}^{+  0.28} $ &  -1.52 &  $    9.97_{ -0.01}^{+  0.14} $ & $    9.75_{  0.00}^{+  0.39} $  \\

  \enddata
\end{deluxetable}

\clearpage
\pagestyle{empty}
\begin{deluxetable}{lrrr}
  \tablecolumns{4} \tablewidth{0pc} \tablecaption{Same as Table~\ref{t6.tab}
    but for fixed-Z from literatures.
    \label{t7.tab}}
  \tablehead{\colhead{ID} &  \colhead{$\rm [Fe/H]_{P10/P11} $} &  \colhead{log $t$} &  \colhead{log $t\rm_{P10/P11}$} \\
    \colhead{} &  \colhead{(dex)} & \colhead{(Gyr)} & \colhead{(Gyr)} }
  \startdata
     B015D-D041 &   -0.02 &  $    8.95_{ -0.95}^{+  1.18} $ & $    7.85_{ -0.15}^{+  0.15} $  \\
    B040-G102 &   -0.02 &  $    8.78_{ -0.22}^{+  0.16} $ & $    7.90_{ -0.15}^{+  0.20} $  \\
    B043-G106 &   -0.02 &  $    8.70_{ -0.18}^{+  0.17} $ & $    7.90_{ -0.15}^{+  0.20} $  \\
    B066-G128 &   -0.02 &  $    8.70_{ -0.31}^{+  0.12} $ & $    7.85_{ -0.15}^{+  0.15} $  \\
    B081-G142 &   -0.02 &  $    9.11_{ -0.32}^{+  0.41} $ & $    8.15_{ -0.15}^{+  0.15} $  \\
   B257D-D073 &   -0.02 &  $    8.85_{ -0.29}^{+  0.46} $ & $    7.90_{ -0.15}^{+  0.20} $  \\
    B318-G042 &   -0.40 &  $    8.85_{ -0.27}^{+  0.12} $ & $    7.85_{ -0.15}^{+  0.15} $  \\
    B321-G046 &    0.65 &  $    8.78_{ -0.16}^{+  0.18} $ & $    8.23_{ -0.15}^{+  0.10} $  \\
    B327-G053 &   -0.40 &  $    8.85_{ -0.27}^{+  0.12} $ & $    7.70_{ -0.10}^{+  0.15} $  \\
    B376-G309 &   -0.02 &  $    8.85_{ -0.27}^{+  0.46} $ & $    8.00_{ -0.15}^{+  0.15} $  \\
    B448-D035 &   -0.02 &  $    9.00_{ -0.09}^{+  0.22} $ & $    7.90_{ -0.15}^{+  0.20} $  \\
    B475-V128 &   -0.40 &  $    9.28_{ -0.13}^{+  0.46} $ & $    8.30_{ -0.20}^{+  0.20} $  \\
         V031 &   -0.70 &  $    9.15_{ -0.11}^{+  0.99} $ & $    8.45_{ -0.15}^{+  0.15} $  \\
   VDB0-B195D &   -0.02 &  $    8.70_{ -0.18}^{+  0.12} $ & $    7.40_{ -0.30}^{+  0.30} $  \\
    B083-G146 &   -0.40 &  $    9.96_{ -0.16}^{+  0.18} $ & $    8.70_{  0.00}^{+  1.44} $  \\
    B222-G277 &   -0.02 &  $   10.11_{ -0.79}^{+  0.03} $ & $    8.60_{  0.00}^{+  1.54} $  \\
    B347-G154 &   -0.40 &  $   10.13_{ -0.33}^{+  0.01} $ & $    8.80_{  0.00}^{+  1.34} $  \\
    B374-G306 &   -0.02 &  $    9.11_{ -0.26}^{+  0.86} $ & $    8.50_{  0.00}^{+  1.64} $  \\
         NB16 &   -0.02 &  $    9.68_{ -0.36}^{+  0.46} $ & $    8.70_{  0.00}^{+  1.44} $  \\
    B049-G112 &   -0.02 &  $    9.26_{ -0.07}^{+  0.72} $ & $    8.45_{ -0.20}^{+  0.20} $  \\
    B367-G292 &   -0.02 &  $    8.95_{ -0.15}^{+  0.36} $ & $    8.30_{ -0.20}^{+  0.20} $  \\
    B458-D049 &   -0.02 &  $    9.15_{ -0.66}^{+  0.97} $ & $    8.50_{ -0.20}^{+  0.20} $  \\
  B521-SK034A &   -0.02 &  $    8.70_{ -0.79}^{+  1.44} $ & $    8.60_{ -0.30}^{+  0.30} $  \\
         M039 &   -0.02 &  $    8.90_{ -0.68}^{+  1.07} $ & $    8.50_{ -0.20}^{+  0.20} $  \\
         M050 &   -0.02 &  $    8.78_{ -0.37}^{+  1.33} $ & $    8.75_{ -0.30}^{+  0.30} $  \\
    B315-G038 &   -0.40 &  $    8.90_{ -0.33}^{+  0.09} $ & $    8.00_{ -0.20}^{+  0.15} $  \\
    B319-G044 &   -0.40 &  $    8.85_{ -0.07}^{+  0.18} $ & $    8.00_{ -0.20}^{+  0.15} $  \\
    B342-G094 &   -0.40 &  $    8.95_{ -0.21}^{+  0.25} $ & $    8.20_{ -0.20}^{+  0.15} $  \\
    B368-G293 &   -0.02 &  $    8.70_{ -0.41}^{+  0.19} $ & $    7.80_{ -0.10}^{+  0.10} $  \\
    B292-G010 &   -1.82 &  $    9.96_{ -0.38}^{+  0.16} $ & $    9.80_{  0.00}^{+  0.34} $  \\
    B337-G068 &   -1.30 &  $   10.00_{ -0.17}^{+  0.12} $ & $    9.65_{  0.00}^{+  0.49} $  \\
    B350-G162 &   -1.52 &  $   10.00_{ -0.35}^{+  0.13} $ & $    9.75_{  0.00}^{+  0.39} $  \\

  \enddata
\end{deluxetable}

\clearpage

\label{lastpage}
\end{document}